\documentclass[10pt,aps,prx,twocolumn,superscriptaddress,nofootinbib]{revtex4-2}
\usepackage{amsmath,amssymb,amsfonts,mathtools,mathrsfs}
\usepackage{amsthm}
\usepackage{bbm}
\usepackage{hyperref}
\usepackage[table,xcdraw]{xcolor}
\usepackage{dsfont}
\usepackage{ifthen}
\usepackage{tkz-euclide} 
\usepackage{pgfplots}
\pgfplotsset{compat=1.18}
\usetikzlibrary{babel, pgfplots.fillbetween}
\pgfdeclarelayer{pre main}
\pgfsetlayers{pre main, main}
\usepackage{pifont}
\usepackage{enumitem}
\usepackage[normalem]{ulem}

\newcommand\doublecheck{\textcolor{black}{\checkmark\kern-0.6em\checkmark}}

\makeatletter

\def\CT@@do@color{%
      \global\let\CT@do@color\relax
            \@tempdima\wd\z@
            \advance\@tempdima\@tempdimb
            \advance\@tempdima\@tempdimc
    \advance\@tempdimb\tabcolsep
    \advance\@tempdimc\tabcolsep
    \advance\@tempdima2\tabcolsep
            \kern-\@tempdimb
            \leaders\vrule
                    \hskip\@tempdima\@plus  1fill
            \kern-\@tempdimc
            \hskip-\wd\z@ \@plus -1fill }
            
\makeatother

\newcommand{\ck}[1]{\cellcolor{lime}\ifthenelse{#1=1}{\!\!\cite{PRT}\!\!}{}\ifthenelse{#1=2}{$\cdot$}{}\ifthenelse{#1=3}{$\circ$}{}\ifthenelse{#1=4}{$\checkmark$}{}\ifthenelse{#1=5}{$\bullet$}{}\ifthenelse{#1=6}{$\mathrm{H}$}{}{}}
\newcommand{\cro}{\cellcolor{red}\ding{55}}
\newcommand{\gro}{\cellcolor{lightgray}}
\newcommand{\unk}{}

\newcommand{\Tr}{\operatorname{Tr}}
\newcommand{\ket}[1]{|#1\rangle}
\newcommand{\bra}[1]{\langle #1|}
\newcommand{\brakett}[1]{\langle#1\rangle}
\newcommand{\braket}[2]{\langle#1|#2\rangle}
\newcommand{\id}{\mathbbm 1}

\DeclareMathOperator{\N}{\mathbb{N}}

\DeclareMathOperator{\R}{\mathbb{R}}
\DeclareMathOperator{\C}{\mathbb{C}}
\DeclareMathOperator{\Hl}{\mathcal{H}}
\DeclareMathOperator{\Image}{Image}
\DeclareMathOperator{\Real}{Re}
\DeclareMathOperator{\Imag}{Im}
\DeclareMathOperator{\vspan}{span}
\DeclareMathOperator{\ord}{ord}
\DeclareMathOperator{\lcm}{lcm}

\DeclareMathOperator{\SL}{SL}
\DeclareMathOperator{\LU}{LU}
\DeclareMathOperator{\GHZ}{GHZ}
\DeclareMathOperator{\LME}{LME}
\DeclareMathOperator{\Cl}{Cl}
\DeclareMathOperator{\orb}{orb}
\DeclareMathOperator{\rank}{rank}
\DeclareMathOperator{\diag}{diag}
\DeclareMathOperator{\tr}{tr}
\DeclareMathOperator{\T}{T}

\newtheorem{theorem}{Theorem}[section]
\newtheorem{prop}[theorem]{Proposition}
\newtheorem{corollary}[theorem]{Corollary}
\newtheorem{lemma}[theorem]{Lemma}
\newtheorem{definition}[theorem]{Definition}
\newtheorem{conj}[theorem]{Conjecture}
\theoremstyle{definition}
\newtheorem{examp}[theorem]{Example}
\newtheorem{rem}[theorem]{Remark}

\newcommand{\ketbra}[2]{\ket{#1}\!\bra{#2}}

\begin{document}

\title{
When can local unitaries generate a complete orthogonal basis?}

\author{Julian Eisfeld}
\email{julian.eisfeld@hhu.de}
\author{Hermann Kampermann}
\author{Nikolai Wyderka}
\affiliation{Institut für Theoretische Physik III, Heinrich-Heine-Universität Düsseldorf, Universitätsstraße 1, 40225 Düsseldorf, Germany}
\date{\today}

\begin{abstract}
The question of local encodability asks: can a given, multipartite pure quantum state be completed to an orthonormal basis using only local unitary rotations? If any such state in a given system has this property, it is called universally locally encodable (ULE), a feature, which has applications in quantum information theory, engineering and math. Despite its easy formulation, the answer to this question depends nontrivially on the local dimensions of the system: While universal local encodability is known for $2\times 2$, $4\times 4$ and $8\times 8$-dimensional systems, it was so-far only conjectured to hold in $3\times 3$, and non-ULE was recently proved for systems of four and more qubits by Pauwels and Gühne. 

We resolve the open conjecture by proving universality for two qutrits. Furthermore, we construct (among others) universal local encodings in dimensions $d\times2$ for $d\ge2$, in dimensions $d\times3$ for $d=3,4$ and $d\ge6$, and in every bipartite system with one or two infinite-dimensional subsystems. For finite-dimensional multipartite systems, we turn a heuristic parameter count into a necessary condition for universality, which excludes universal local encodability in infinitely many cases. Finally, we give universal local encodability an operational interpretation by showing that non-ULE is equivalent to whether general measurements yield an advantage over projectively simulable ones in discrimination tasks where unknown, random local unitary rotations have to be detected.
\end{abstract}

\maketitle

\section{Introduction}

Entanglement is a central feature of quantum mechanics and constitutes a resource for many tasks in quantum communication, quantum networks and distributed quantum computing. While the entanglement of quantum states has been studied extensively, it can also enter via quantum measurements, since the basis states of a joint projective measurement may themselves be entangled. The most prominent example is the Bell-state measurement, which is an essential ingredient for quantum teleportation \cite{Bennett1993Teleportation}, dense coding \cite{Bennett1992DenseCoding} and entanglement swapping \cite{Zukowski1993EntanglementSwapping}. In these protocols, entanglement therefore enters both through the states shared between the parties and through the measurement used to process these states.

A characteristic property of the Bell basis is that all four Bell states are connected by local unitary transformations. But is this a special feature of the Bell basis, or a general feature of quantum states?
This leads to the following general question, known as the problem of local encodability~\cite{Tanaka}: Given a pure state $\ket{\psi}$ shared between $n$ parties, does its local-unitary orbit contain an orthonormal basis? More precisely, for local Hilbert-space dimensions $d_1,\ldots,d_n$ and total dimension $D=\prod_k d_k$, are there $D$ local unitaries
\begin{equation*}
V_a= U_1^{(a)} \otimes U_2^{(a)} \otimes \ldots \otimes U_n^{(a)},
\end{equation*}
such that the states $\ket{\psi_a}=V_a\ket{\psi}$ form an orthonormal basis, i.e., satisfy $\brakett{\psi_a|\psi_b}=\delta_{ab}$? If such local unitaries exist, the state is called locally encodable. If all states in a given multipartite Hilbert space are locally encodable, we call it universally locally encodable (ULE). Operationally, the $D$ basis labels correspond to $D$ perfectly distinguishable classical messages that can be encoded using only coordinated local unitary transformations and subsequently read out by a global measurement. Equivalently, the state can be associated with a complete projective measurement in which every basis state has the same local-unitary invariants and therefore the same entanglement as the original state.

For the case of $n=2$, the same question has a direct formulation outside of quantum information. Since the coefficients of a bipartite quantum state can be interpreted as a $d_A \times d_B$-dimensional matrix, local encodability of an arbitrary state $\ket{\psi}$ is equivalent to asking whether the two-sided unitary orbit of an arbitrary matrix contains a Hilbert-Schmidt orthonormal matrix basis. Such a basis is a complete orthogonal set in which every matrix has the same rank, the same singular values and the same Schatten norms. This connects the problem to structured orthogonal systems and tight frames in mathematics and signal processing \cite{Strohmer2003GrassmannianFrames}, and to matrix sensing.
In all of these formulations, the central question is the same: Can one matrix with a fixed singular-value spectrum be rotated into sufficiently many mutually orthogonal directions to span the entire matrix space?

Although the problem has a simple formulation, its solution is not obvious, as the $D$ locally rotated states have to satisfy $D(D-1)/2$ complex orthogonality constraints simultaneously.
Previous work on local encodability focused mainly on specific families of multipartite states. In Ref.~\cite{Tanaka}, local encodings were constructed for stabilizer states and states generated through pseudo-Clifford operations. Additionally, equientangled bases that continuously interpolate between product and maximally entangled bases were constructed for bipartite systems of arbitrary local dimension~\cite{Karimipour2006Equientangled,Gheorghiu2010Construction}. While these results establish the existence of bases with continuously tunable entanglement, they do not determine whether an arbitrary prescribed state can be extended to a basis. More recently, Ref.~\cite{DelSanto2024IsoEntangled} provided a complete classification of equientangled bases for two qubits, while only partial results were obtained in higher dimensions. In Ref.~\cite{PRT}, local encodability for all states was proven for bipartite systems of equal local dimension $d_A = d_B \in \{2,4,8\}$ (where we use $d_1=d_A$ and $d_2 = d_B$ in bipartite systems. Throughout the paper, we assume $d_A \geq d_B$), and numerical evidence was provided for systems of two qutrits and three qubits, together with the conjecture that there exist four qubit states which are not locally encodable. This  conjecture was recently proven in Ref.~\cite{Guehne}, where a specific counter-example was constructed. This implies, via a result from Ref.~\cite{PRT}, that there exist such states for any number of four or more qubits. Nevertheless, an analytical solution for two qutrits, constructions for unequal or infinite local dimensions, and general criteria for the (non-)existence of local encodability remained unknown. What is more, despite of its numerous links to different fields of math, we so far lacked an appealing operational interpretation in the field of quantum information in terms of a proper advantage in some task if universal local encodability was known to hold. 

In this paper we find this operational interpretation in terms of the existence of optimal projective measurement in some local invariant estimation task. Additionally, we provide explicit constructions showing universal local encodability for infinitely many bipartite system dimensions, including $d_A \times 2$ for all $d_A\geq 2$, $d_A \times 3$ for all $d_A \geq 3$ except of $d_A= 5$ and $\infty \times d_B$ for all $d_B$, including $d_B=\infty$. Furthermore, we generalize non-existence arguments to many multipartite systems, provided their local dimensions violate a naive degree-of-freedom counting inequality. The findings are summarized in Tables~\ref{tab:constructions} and~\ref{tab:constructions2}. The paper is organized as follows. In Section~\ref{sec:elementary}, we introduce the notation and basics and provide a first useful lifting lemma that allows to extend universal local encodability for dimension pairs $d_A \times d_B$ and $d_{A'} \times d_B$ to $(d_A+d_{A'})\times d_B$ if $d_A, d_{A'} \geq d_B$. In Section~\ref{sec:hurwitz}, we show that for each choice of $d_B$, there exists a finite $d_A \geq d_B$ such that $d_A\times d_B$ is universal local encodable using a Hurwitz-Radon construction. Using the lifting lemma, this establishes ULE in $\infty \times d_B$ systems for all $d_B$. Section~\ref{sec:group} introduces a family of group based constructions that we exploit to show ULE in dimensions $3\times 3$ which extends, together with the Hurwitz-Radon construction and the lifting lemma to all $d_A \times 3$ except of $d_A=5$, and we show that $5 \times 3$ is not tractable using this approach. Finally, in Section~\ref{sec:nogo}, we turn to the multipartite case and generalize the no-go theorem of ULE for four-qubit systems to any system of local dimensions violating a certain inequality.

Finally, in Section~\ref{sec:application}, we give universal local encodability an operational interpretation by showing that ULE is equivalent to the existence of an optimal projective measurement in a class of locally invariant estimation tasks.
\onecolumngrid

\begin{table}[t]
\begin{tabular}{c||c|c|c|c|c|c|c|c||c||c|}
$d_{B}\setminus d_{A}$ & $2$ & $3$ & $4$ & $5$ & $6$ & $7$ & $8$ & $9$ & $\cdots$& $\infty$\tabularnewline
\hline 
\hline 
$2$ & \ck1 /$\mathrm{H}/\doublecheck $& \ck2 /$\checkmark$ & \ck2 /$\doublecheck$& \ck2 /$\checkmark$& \ck2 /$\doublecheck$& \ck2 /$\checkmark$& \ck2 /$\doublecheck$& \ck2/$\checkmark$ & \ck2 / $\checkmark$ ($\,2|d_A$: $\doublecheck$)& \ck3\tabularnewline
\hline 
$3$ & \gro & \ck4 & \ck4 / \ck6 & \unk & \ck5 & \ck5 & \ck5 & \ck5 & \ck5& \ck3\tabularnewline
\hline 
$4$ & \gro & \gro & \ck1  / $\mathrm{H} $& \ck2 & \ck2 & \ck2 & \ck2 & \ck2 & \ck2& \ck3\tabularnewline
\hline 
$5$ & \gro & \gro & \gro & \unk & \unk & \unk & \ck6 & \unk & \cellcolor{yellow}$(\,8|d_A\!: \; \bullet)$ &\ck3\tabularnewline
\hline 
$6$ & \gro & \gro & \gro & \gro & \unk & \unk & \ck6 & \unk & \cellcolor{yellow}$(\,8|d_A\!: \; \bullet)$ & \ck3\tabularnewline
\hline 
$7$ & \gro & \gro & \gro & \gro & \gro & \unk & \ck6 & \unk & \cellcolor{yellow}$(\,8|d_A\!: \; \bullet)$ & \ck3\tabularnewline
\hline 
$8$ & \gro & \gro & \gro & \gro & \gro & \gro & \ck1  / $\mathrm{H} $& \ck2 & \ck2 & \ck3\tabularnewline
\hline 
$9$ & \gro & \gro & \gro & \gro & \gro & \gro & \gro & \unk & \cellcolor{yellow}$(\,16|d_A\!: \; \mathrm{H}/\bullet)$ & \ck3\tabularnewline
\hline
\hline
$\cdots$ & \gro & \gro & \gro & \gro & \gro & \gro & \gro & \gro & \cellcolor{yellow}$\mathrm{H}/\bullet$ & \ck3\tabularnewline
\hline
\hline

$\infty$ & \gro & \gro & \gro & \gro & \gro & \gro & \gro & \gro & \gro &\cellcolor{lime} $\mathrm{H}_\infty$
\tabularnewline
\hline
\end{tabular}
    \caption{List of results in bipartite systems with dimensions $d_A \times d_B$ (with $d_A \geq d_B$), together with their current status regarding provable  universal local encodability. A white cell denotes the status of unknown, a green cell the status of provably true, and a yellow cell the status of provably true for a subset of dimensions. For example, for $d_B=5$ there are only constructions available if $d_A$ is a multiple of $8$ (denoted by $8|d_A$). There are several different methods of constructions, differentiated by the symbols used. A citation means, that the result was already known and proven in the cited paper. A small dot ($\cdot$) means, that the result follows from a simple extension (Theorem~\ref{th:dx2_ext}). A large dot ($\bullet$) means, that the result follows from a lifting lemma, while a circle ($\circ$) refers to its modification for $\infty \times d_B$ (Lemma~\ref{lem:enc_poss_multi} for both). A check mark ($\checkmark$) denotes the construction via the group based approach (Theorems~\ref{th:3x3} and~\ref{alt 4x3 theorem} as well as Proposition~\ref{dx2}), while a double check mark ($\doublecheck$) denotes the existence of multiple different such constructions. $\mathrm{H}$ refers to a construction using Hurwitz-Radon matrices (Proposition~\ref{prop:Radon}), $\mathrm{H}_\infty$ to its heavy modification needed for $\infty \times \infty$ (see Appendix Section~\ref{HR oo}).}
    \label{tab:constructions}
\end{table}
\begin{table}[t]
\begin{tabular}{c||c|c|c|c|c|c|c||c||c|}
$n\setminus d$ & $2$ & $3$ & $4$ & $5$ & $6$ & $7$ & $8$ & $\cdots$& $\infty$\tabularnewline
\hline 
\hline 
$2$ & \ck1 / $\mathrm{H}/\doublecheck$  & \ck4 & \ck1 / $\mathrm{H}$ & & & & \ck1 / $\mathrm{H}$ &  & \cellcolor{lime} $\mathrm{H}_\infty$\tabularnewline
\hline 
$3$ &  & \cro & \cro  & \cro & \cro & \cro & \cro & \cro & \tabularnewline
\hline 
$4$ & \cellcolor{red} \cite{Guehne} & \cro & \cro & \cro & \cro & \cro & \cro & \cro & \tabularnewline
\hline 
$\geq5$ & \cellcolor{red} \cite{Guehne} \& \cite{PRT}& \cro & \cro & \cro & \cro & \cro & \cro & \cro &\tabularnewline
\hline 
\end{tabular}
    \caption{List of multipartite results for different numbers of parties $n$ and local dimension $d$ (under the assumption that they are all equal), together with their current status regarding provable universal local encodability. A red cell denotes the status of provably false, a green cell the status of provably true and a white cell the status of unknown. There are several different methods of proof, differentiated by the symbols used. Those already used in Table~\ref{tab:constructions} are explained there. Additionally: A double citation refers to the claim resulting from combining statements from the two papers and a cross (\ding{55}) denotes a violation of Inequality (\ref{eq:dim_ineq}), which yields the impossibility of universal local encodability by Theorem~\ref{th:impossible_main}. }
    \label{tab:constructions2}
\end{table}
\newpage

\twocolumngrid

\section{Elementary Observations and Results}\label{sec:elementary}
\subsection{Notation and Definitions}
For the entire document let $n\in \mathbb{N}_{\geq 2}$ denote the number of parties. For $k=1,...,n$ let $\Hl_k$ be a (separable) complex Hilbert space of dimension $d_k \in \mathbb{N}_{\geq 2} \cup \{\infty\}$ and set $\Hl = \bigotimes\limits_{k=1}^n \Hl_k$. Denote its dimension by $D$.

Let us define the objects of interest:
\begin{itemize}[leftmargin=*]
\item The \textbf{group of local unitary operators} on $\Hl$ is given by
\begin{equation*}
    \quad \LU(d_1,...,d_n)\! =\! \{U_1\otimes...\otimes U_n  | \forall k : U_k \in \mathrm{U}(d_k)  \}
\end{equation*}
or simply by $\LU$ if the dimensions are clear. 
\item Two states $\ket{\psi}, \ket{\varphi} \in \Hl$ are called \textbf{$\LU$-equivalent} (denoted by $\ket{\psi} \stackrel{\LU}{\sim} \ket{\varphi}$), if there is a $V \in \LU$, such that $\ket{\psi} = V\ket{\varphi}$.

\item A set of states $S \subseteq \Hl$ is called \textbf{$\LU$-universal}, if every $\ket{\psi}\in \Hl$ is $\LU$-equivalent to a state in $S$. 

\item The \textbf{$\LU$-orbit} of $\ket{\psi}$ is given 
\begin{equation*}
    \orb_{\LU}(\ket{\psi}) = \left\{\ket{\varphi} \in \Hl : \ket{\varphi} \stackrel{\LU}{\sim} \ket{\psi}\right\}.
\end{equation*}

\item A state $\ket{\psi}$ is called \textbf{locally encodable}, if $\orb_{\LU}(\ket{\psi})$ contains a complete orthonormal basis. We call the local unitaries that map $\ket{\psi}$ to a basis the \textbf{local encoders} of $\ket{\psi}$.

\item We call $\mathcal{H}$ \textbf{universal locally encodable} (ULE), if all states $\ket{\psi} \in \mathcal{H}$ are locally encodable.
\end{itemize}

Note that, since all complex separable Hilbert spaces of the same dimension are isomorphic to each other, the question of universal local encodability only depends on $n$ and $d_1,...,d_N$. Allowing $d_j = \infty$ corresponds to $\mathcal{H}_j \cong \ell^2$, which can physically correspond to the Hilbert space associated to a harmonic oscillator.

\subsection{Heuristics on universal local encodability}\label{subsec:heu}
Let us start by providing a heuristic argument as to when universal local encodability is expected to hold (provided $d_1,...,d_n < \infty$).

Fix a state $\ket{\psi} \in \mathcal{H}$, and consider sets of vectors $\{V_j \ket{\psi}, j=1,...,D\}$ with $V_j \in \LU(d_1,...,d_n)$. They span an orthonormal basis if and only if $\braket{V_j \psi}{V_k\psi} = \delta_{jk}$. There are $D-1$ free matrices (since the first matrix can always be chosen to be $\id$), each including $n$ unitaries which each have $d_j^2-1, j \in \{1,...,n\}$ real degrees of freedom (after accounting for an irrelevant phase in each of them). The total number of degrees of freedom is thus $(D-1)\sum_j (d_j^2-1)$ while the number of real constraints due to orthogonality between the $D$ vectors is $D(D-1)$. Thus one would heuristically expect universal local encodability if and only if
\begin{equation}\label{eq:dim_ineq}
    \sum\limits_{j=1}^n d_j^2 \stackrel{!}{\geq} \prod\limits_{j=1}^n d_j+n,
\end{equation}
Notably universal local encodability should be possible for any amount of parties, as long as their dimensions are suitably chosen (take for example ${d_1 = ... = d_{n-1} =2}$ and $d_n = 2^n$).
Furthermore, for $n=2$ it is expected to be possible for arbitrary local dimensions. For ${d=d_1=...=d_n \in \mathbb{N}_{\geq 2}}$ $(n\geq 3)$ the resulting inequality is only fulfilled for $n=3$, $d=2$.

We take this heuristic criterion as a guideline for the following chapters: It predicts ULE in any bipartite system, and we show in Sections~\ref{sec:hurwitz} and~\ref{sec:group} that it indeed holds for many choices of local dimensions. Contrary, the heuristic predicts that in many multipartite systems, ULE should not hold. In Section~\ref{sec:nogo}, we show that indeed, failing of inequality (\ref{eq:dim_ineq}) implies that universal local encodability is impossible.

\subsection{Infinite Dimensions and Lifts}\label{subsec:inf_dim}
Before showing universal local encodability in particular settings, let us start by noting that in the  bipartite case, existing results of universal local encodability can be lifted to higher dimensions, including infinite dimensional systems:
\begin{lemma}[Lifting]\label{lem:enc_poss_multi}
    Let $d_A,d_A',d_B \in \N_{\geq 2}$ with $d_A,d_A' \geq d_B$.
    If universal local encodability holds for $d_A \times d_B$ and $d_A' \times d_B$-dimensional systems, then it holds for $(d_A+d_A')\times d_B$-dimensional systems as well, in particular for $\kappa d_A \times d_B$ for all $\kappa \in \N \cup \{\infty\}$.
\end{lemma}
\begin{figure}
    \centering
    \includegraphics[width=1.0\columnwidth]{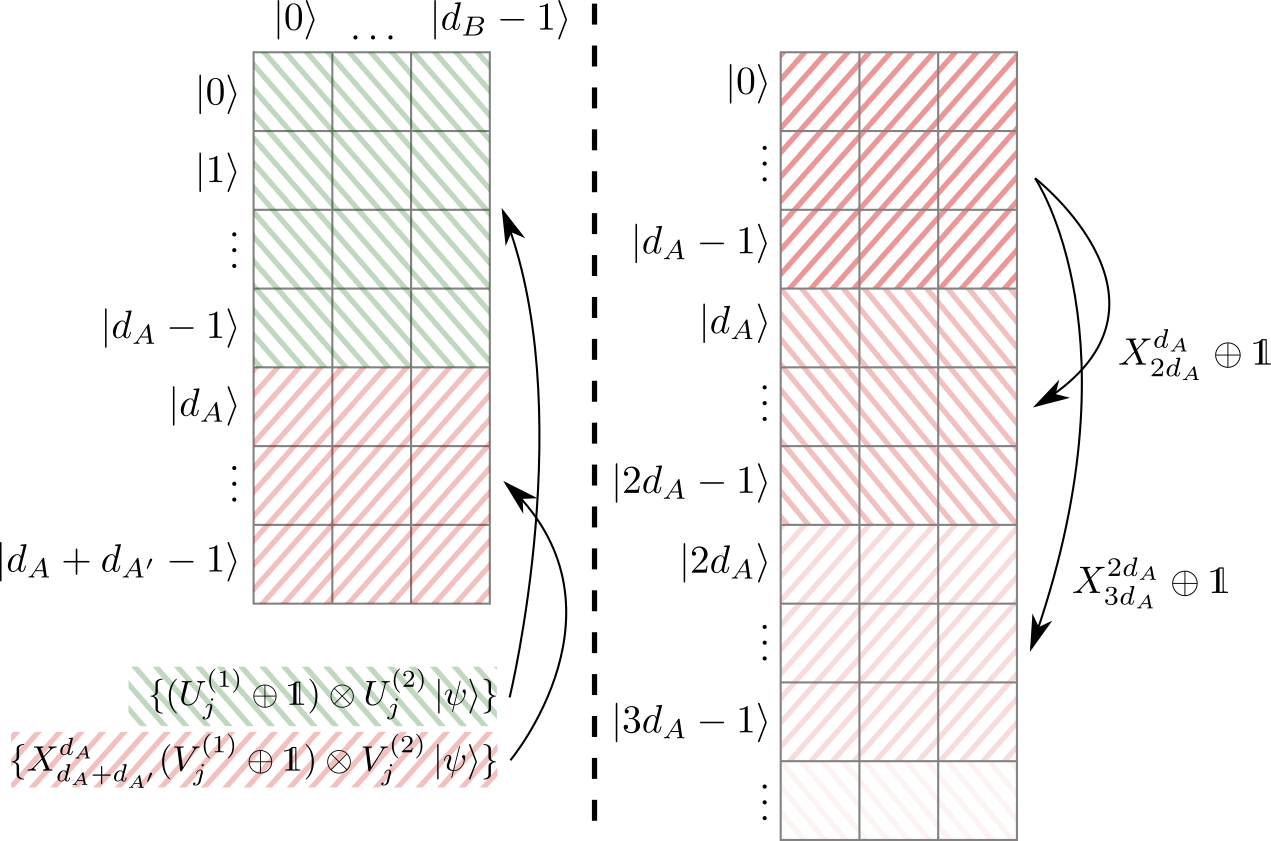}
    \caption{Visualization of the lifting lemma. Left: An encoding for $d_A\times d_B$ and $d_{A'} \times d_B$ with $d_A, d_{A'} \geq d_B$ can be lifted to an encoding of $(d_A + d_{A'})\times d_B$ by shifting the encoded state with $X_{d_A+d_{A'}}^{d_A} \otimes \id$. Right: Any encoding for $d_A \times d_B$ with $d_A \geq d_B$ can be lifted to an encoding of $\infty \times d_B$ via the infinite dimensional unitary operators $X_{(\kappa+1)d_A}^{d_A\kappa} \oplus \id$ with $\kappa \in \N$.}
    \label{fig:lifting}
\end{figure}
\begin{proof}
    Write $\ket{\psi} \in \C^{d_A+d_A'} \otimes \C^{d_B}=(\C^{d_A} \oplus \C^{d_A'}) \otimes \C^{d_B}$ in its Schmidt decomposition $\ket{\psi} = \sum\limits_{k=0}^{d_B-1} c_k \ket{kk}$. 
    Let $(U_j^{(1)} \otimes U_j^{(2)})_{j}$ ($(V_j^{(1)} \otimes V_j^{(2)})_{j}$) be the local encoders of the Schmidt diagonal state with the same Schmidt coefficients in $\C^{d_A} \otimes \C^{d_B}$ ($\C^{d_A'} \otimes \C^{d_B}$). 
    
    Then $\{(U_j^{(1)}\oplus \id)\otimes U_j^{(2)} \ket{\psi}\}$ is a basis for\\ $\vspan\{\ket{k,m} : k=0,...,d_A-1,m =0,...,d_B-1\}$ and $\{(X^{d_A}_{d_A+d_A'}\otimes \id)((V_j^{(1)} \oplus \id )\otimes V_j^{(2)}) \ket{\psi}\}$ is a basis for $\vspan\{\ket{k,m} : k=d_A,...,d_A+d_A'-1,m =0,...,d_B-1\}$, where
    $X_{d_A+d_A'}\ket{j} = \ket{j+1 \, \mathrm{mod} \; d_A+d_A'}$ is the shift matrix. The procedure is visualized in Fig.~\ref{fig:lifting}.

    The ULE claim for $\kappa d_A \times d_B$ now follows from induction for $\kappa<\infty$.

    For the case of $d_A = \infty$, meaning $\mathcal{H} = \ell^2 \otimes \C^{d_B}$, the construction has to be modified. Indeed,
    $\{ \ket{k\!+\!\kappa d_A,m}\!:\! k = 0,...,d_A-1, m = 0,...,d_B-1,  \kappa\! \in \!\N_0\}$ is an orthonormal basis for $\mathcal{H}$, where successive shifts by $d_A$ in $\mathcal{H}_A$ yields new orthogonal vectors from the first $d_Ad_B$ ones.
    But the natural generalization of $X_d$ to $\ell^2$ given by $X_\infty\ket{j} = \ket{j+1}$ is not surjective ($\ket{0}$ is not in its image), and thus not unitary. This can be fixed by considering unitaries of the form $X_d \oplus \mathbbm{1}$ and applying $X_{(\kappa+1)d_A}^{d_A\kappa} \oplus \mathbbm{1}$ for $\kappa\in \N$ to repeatedly get $d_Ad_B$ new orthogonal vectors, that end up spanning the whole space (see Fig.~\ref{fig:lifting}).
\end{proof}
This allows to lift the constructions of Ref.~\cite{PRT} for $2\times 2$, $4 \times 4$ and $8\times 8$ to $2\kappa\times2$, $4\kappa\times 4$ and $8\kappa\times8$ for $\kappa \in \mathbb{N}$, respectively. A similar construction especially adapted to the construction from Ref.~\cite{PRT} can fill the gaps as well:
\begin{theorem}\label{th:dx2_ext}
    Universal local encodability holds $d_A \times d_B$ for $d_A \in \N_{\geq d_B} \cup \{\infty\}$ and $d_B \in \{2,4,8\}$.
\end{theorem}
\begin{proof}
    Fix $d_B \in \{2,4,8\}$ and take any $\ket{\psi} \in \mathcal{H}_A \otimes \mathcal{H}_B$ with $d_A \geq d_B$. The construction in Ref.~\cite{PRT} provides $d_B$ unitaries that map $\ket{\psi}\stackrel{\LU}{\sim} \sum\limits_{k=0}^{d_B-1} c_k \ket{kk}$ to a basis of $\vspan\{ \ket{kk} : k=0,...,d_B-1\}$ (note that this is only possible for $d_B = 2,4$ or $8$
    if those $d_B$ unitaries are required to be state independent, see Appendix Proposition~\ref{sym limit}). Applying $(X_{d_A}^j \otimes \id)$ to this set for $j \in \{0,...,d_A-1\}$ yields a complete orthonormal basis of $\mathcal{H}_A \otimes \mathcal{H}_B$.
\end{proof}

\section{The Hurwitz-Radon Approach}\label{sec:hurwitz}
Let us now turn to the case of general bipartite systems. We show that for each fixed $d_B \in \N$, there is always a $d_A \in \N$, such that $d_A \times d_B$-dimensional systems are universally locally encodable. In particular, using the lifting lemma, this implies that $\infty \times d_B$ for all $d_B \in \N_{\geq 2}$ is ULE.

Our construction hinges on the existence of Hurwitz-Radon matrices, a set of real $d_A\times d_A$ matrices $(A_j)_j$ satisfying $A_j^{\T} A_k + A_k^{\T} A_j = 2\delta_{jk} \id$. For any fixed $d_A$, it is known that there exist $\varrho(d_A)$ matrices, the so called Radon-number given by $\varrho(d_A) = 8 \alpha +2^\beta$, where $d_A=2^{4\alpha +\beta} d'$ (with $d'$ odd, $\beta \in \{0,1,2,3\}, \alpha \in \N_0$), see~\cite{Hurwitz, radon1922lineare}. For a more modern perspective see~\cite{Eckmann} Sections 1 \& 4.

The following is proven in Appendix~\ref{Ap HR}:

\begin{prop}\label{prop:Radon}
    Let $\mathcal{H}=\mathbb{C}^{d_A} \otimes \mathbb{C}^{d_B}$. If $\varrho(d_A) \geq d_B$, where $\varrho$ denotes the Radon number, then $\mathcal{H}$ is universally locally encodable. Furthermore,  for all $d_B \in \N$, $\infty \times d_B$-dimensional systems are universally locally encodable.
\end{prop}
For a fixed $d_B$, the smallest $d_A$ that fulfills $\varrho(d_A) \geq d_B$ is given by OEIS sequence A034583, see~\cite{oeis}. $d_A = 2^{\lceil d_B/2 \rceil}$ always works, but is at times too large by a factor of $2$.

Of particular interest are bipartite systems with $d_A=d_B=\infty$ (for instance describing two coupled harmonic oscillators). However, there are multiple issues in trying to lift above result to $d_B=\infty$. It is apriori not even clear, if one should expect universal local encodability to hold, since the heuristic degree-of-freedom-counting argument is inconclusive. Furthermore, in this case, $\LU$-equivalence is no longer fully governed by the set of Schmidt coefficients, and completeness does not automatically follow from algebraically verifying that vectors in a large enough set are mutually orthogonal. Additionally, there are some nontrivial questions of convergence. Despite all of this, we can construct a countable sequence of Hurwitz-Radon matrices with some additional properties which can be exploited to show ULE for such systems: 
\begin{theorem}[Appendix Theorem~\ref{oo x oo enc}]\label{th:oo_x_oo_main}
    Universal local encodability holds true for two harmonic oscillators (i.e., $d_A =d_B=\infty$). More precisely: Given a normalized element $\ket{\psi} \in \ell^2 \otimes \ell^2$ (or any separable infinite dimensional Hilbert space $\mathcal{H} = \mathcal{H}_1 \otimes \mathcal{H}_2$), there is a family of local unitary operators $(U_j=U^{(A)}_j \otimes U^{(B)}_j)_{j\in \N}$, such that $\{U_j \ket{\psi} : j\in \N\}$ is a complete orthonormal basis.
\end{theorem}

\section{The Group Approach}\label{sec:group}

Showing universal local encodability for a given system is usually a hard task, as one has to construct local encoders for any possible state. However, there is a more structured approach to this problem: Instead of starting with a given state and constructing local encoders, we can in many cases find a group with $D$ elements which automatically leads to local encodability for large classes of states. The remaining problem is then to show that these classes are $\LU$-universal, i.e., that every state is $\LU$-equivalent to some state in that class. 

In particular, we can show the following:
\begin{prop}[Group encodings]\label{prop:group}
    Assume there exists a subgroup 
    $ {G \subseteq \LU(d_1,...,d_n)}$ with the following properties:
    
    (i) $G$ is abelian.
    
    (ii) Every element of $G$ except $\id$ is traceless.
    
    (iii) The number of elements in $G$ is $\#G = D$.\\
    There exists a unique common orthonormal eigenbasis of all $g \in G$, denoted by $\{\ket{\psi_j} : j=0,...,D-1\}$. For every choice of $\theta \in \R^{D}$ the state
    \begin{equation}\label{eq:max_coherent}
        \ket{\psi(\theta)} = \frac{1}{\sqrt{D}} \sum_{j=0}^{D-1} e^{i\theta_j} \ket{\psi_j},
    \end{equation}
    as well as any state that is $\LU$-equivalent to it, is locally encodable.
\end{prop}
\begin{proof}
    The existence of a common eigenbasis follows from (i), its uniqueness is proven in Appendix Lemma~\ref{joint eig}.
    It is sufficient to show that $\left\{ g \ket{\psi(\theta)} : g \in G\right\}$ is an orthonormal basis of $\Hl$. 
    
    $G$ consists of $D$ operators, all of them being local unitaries, so orthogonality is the only concern. Since $G$ is a group of unitaries, for any two elements $g_1,g_2 \in G$ we have to show that
    \begin{equation*}
        \braket{g_1\psi(\theta)}{g_2 \psi(\theta)} = \braket{\psi(\theta)}{g^{-1}_1 g_2\psi(\theta)} \stackrel{!}{=} \delta_{g^{-1}_1g_2, \mathbbm{1}}.
    \end{equation*}
    Denote the $k$-th eigenvalue of $g=g_1^{-1}g_2$ as $\lambda_k^{(g)}$ to write
    \begin{equation*}
    \begin{split}
        \braket{\psi(\theta)}{g\psi(\theta)} &= \frac{1}{D} \sum_{j,k = 0}^{D-1} e^{i \theta_j-\theta_k} \braket{\psi_j}{g\psi_k} \\
        &= \frac{1}{D} \sum_{j,k = 0}^{D-1} e^{i \theta_j-\theta_k} \lambda_k^{(g)} \underbrace{\braket{\psi_j}{\psi_k}}_{=\delta_{jk}}\\
        & = \frac{1}{D} \sum_{k = 0}^{D-1} \lambda_k^{(g)} = \frac{\Tr(g)}{D}  = \delta_{g, \mathbbm{1}}
    \end{split}
    \end{equation*}
    since $\mathbbm{1}$ has trace $D$ and all other elements of $G$ are traceless.
\end{proof}
States of the Form (\ref{eq:max_coherent}) are called \textbf{maximally coherent} in the basis $\{\ket{\psi_j} : j=0,...,D-1\}$. The above construction shows that universal local encodability is possible, if there exists a group $G$, such that the set of states of the form (\ref{eq:max_coherent}) is $\LU$-universal. Note, that the maximally coherent states only depend on the basis and no longer on the group.

The existence of such groups (especially in the bipartite case) is explored in Appendix Section~\ref{Ap group}. There we show the following:
\begin{prop}\label{prop:group_summary} The following holds:

    (a) For any $n,d_1,...,d_n \in \N_{\geq 2}$, there always exists a group fulfilling Proposition~\ref{prop:group}, whose eigenbasis is the product basis. 
    
    (b) If $d_1=...=d_n=2$, the maximally coherent states from (a) are the locally maximally entanglable states, which are known to be locally encodable (see~\cite{Kraus}).
    
    (c) The set of maximally coherent states from (a) can fail to be $\LU$-universal, including for $d\times 3$ (with $d\geq 3$ odd) or for $n \geq 3$ qubits.
    
    (d) For $n=2$, there exists a group with common eigenbasis differing from the product basis, if and only if $\gcd(d_A,d_B) > 1$.
    
    (e) For $d_A=d_B=d$, there exists a group whose eigenbasis is the Bell basis with basis elements
    \begin{equation*}
        \ket{\psi_{a,b}} =(Z^a \otimes X^b)\ket{\phi^+}
    \end{equation*}
    where $\ket{\phi^+} = \frac{1}{\sqrt{d}} \sum\limits_{\ell = 0}^{d-1} \ket{\ell \ell}$, is a generalized Bell state.
\end{prop}
\begin{proof}
    (a) See Example~\ref{prod ex}, (b) \& (c) see Remark~\ref{LME rem}, (d) \& (e) See Proposition~\ref{gcd prop n=2} and Example~\ref{group const}.
\end{proof}

In order to use the group construction to show ULE in a certain system, we have to show  LU-universality of the corresponding maximally coherent states. In the case of bipartite systems, this translates into showing that every possible Schmidt coefficient vector occurs at least once. These vectors are uniquely recoverable from the trace powers of the matrix $B(\theta) \coloneqq A(\theta)^\dagger A(\theta)$, where $A(\theta)$ is the coefficient matrix of $\ket{\psi(\theta)}$ in the computational basis. More precisely, setting
\begin{equation*}
    \tau (B) \coloneqq \left(\Tr(B^2),\Tr(B^3),...,\Tr(B^{d_B})\right)
\end{equation*}
and
\begin{equation*}
    \mathcal{T}_d \coloneqq \left\{ \left( \,\sum_{j=1}^d \lambda_j^2,...,\sum_{j=1}^d \lambda_j^d\right) : \lambda_j \geq 0, \sum_{j=1}^d \lambda_j = 1\right\},
\end{equation*}
then $\tau (B(\theta))$ should span the entirety of $\mathcal{T}_{d_B}$.

By interpolation arguments, it is even sufficient to only find encoders such that $\tau (B(\theta))$ covers $\partial \mathcal{T}_{d_B}$, as long as this happens in a certain continuous way, which is discussed in Appendix Section~\ref{top sec}.

This can be directly used to rediscover the fact that local encodability holds for $d\times 2$-dimensional system (see Appendix Proposition~\ref{dx2}).

Furthermore, we can use this to show universal local encodability for $3\times 3$ by finding a continuous, periodic family of parameters $\theta$ that trace exactly the boundary of the two-dimensional set $\mathcal T_3$ (see Appendix Section~\ref{3x3}). Combining this with the $4\times 3$ result from Proposition~\ref{prop:Radon} (or the alternative construction outlined in Appendix Section~\ref{4x3 app}), the lift in Lemma~\ref{lem:enc_poss_multi} and the $d\times 2$ result from Theorem~\ref{dx2} yields:
\begin{theorem}\label{th:3x3}
    All bipartite systems with dimensions $d \times 3$ are universally locally encodable for $d \in \N_{\geq 2}$, $d \neq 5$.
\end{theorem}
This explicitly solves half of Conjecture I in Ref.~\cite{PRT}.
Furthermore, we numerically checked for the existence of arbitrary Schmidt coefficient vectors in the non-product basis common eigenbasis in non-coprime dimension pairs and could always find a corresponding maximally coherent state. This motivates us to phrase the following conjecture:

\begin{conj}\label{conj:group}
    Let $\gcd(d_A,d_B) > 1$. Then the set of states of the Form (\ref{max co G}), the maximally coherent states in the eigenbasis of the group from Proposition~\ref{prop:group_summary}(d), is $\LU$-universal. This would imply that universal local encodability holds for any bipartite system with $\gcd(d_A,d_B) > 1$, including $d\times d$ for all $d \in \N_{\geq 2}$.
\end{conj}
Note that for $d \times d$ systems, there always is a group with a maximally entangled common eigenbasis, and our conjecture would imply that every $d \times d$-dimensional bipartite state is LU-equivalent to some maximally coherent states in the $d$-dimensional Bell basis.

Interestingly enough, the missing case of $5\times 3$ in Theorem~\ref{th:3x3} cannot be proven using such a group approach.
This follows from combining Proposition~\ref{prop:group_summary}(c) and (d).

We summarize all of our constructions for the bipartite case in Tab.~\ref{tab:constructions}.

\section{No-go Theorems}\label{sec:nogo}
Let us now turn our focus onto multipartite systems. The results of Pauwels and Gühne in Ref.~\cite{Guehne} provide a simple sufficient condition for impossibility of universal local encodability.
Though only explicitly stated for systems of $n$ qubits, it can be generalized. First, we introduce the main ingredients for the proof in Ref.~\cite{Guehne}: 
\begin{definition} 
     Let $\{T_k^{(d)} : k=1,...,d^2-1\}$ be a basis for $i\mathfrak{su}(d)$ (i.e. a set of generators for $SU(d)$ following the physics convention). Let 
    \begin{align*}
    \mathcal{G} &= \bigcup\limits_{j=1}^n \mathcal{G}_j \text{ with }\\
        \mathcal{G}_j &= \{ \mathbbm{1}^{j-1} \otimes T_k^{(d_j)} \otimes \mathbbm{1}^{n-j} : k \in \{1,...,d_j^2-1\}\}
    \end{align*}
     be the set of local generators. A state $\ket{\psi} $ is called \textbf{one-uniform} if and only if $\braket{\psi}{T|\psi} = 0$ for all $T \in \mathcal{G}$. It is called \textbf{regular}, if the map
    \begin{equation*}
        \R^{\#\mathcal{G}} \to \C^{D}, \quad v \mapsto \left(\sum_{T \in \mathcal{G}} v_{T} T\right) \ket{\psi}
    \end{equation*}
    is injective.
\end{definition}
The proof in Ref.~\cite{Guehne} builds on two things: the violation of a dimensional inequality and the nonemptiness of a certain set ($\mathcal{Z}^{\text{reg}}$ in their notation), which is equivalent to the existence of a one-uniform regular state. Such a state is then explicitly constructed for four qubits. 

A direct generalization of the impossibility theorem can be done by replacing all occurrences of Pauli matrices with the corresponding local generators, which results in the following criterion:
\begin{prop}\label{prop:criteria}
    If $\mathcal{H} = \bigotimes_{k=1}^n \mathbb{C}^{d_k}$ contains a one-uniform, regular state $\ket{\psi}$ and
    \begin{equation}\label{ineq 2}
        \sum_{j=1}^n d_j^2 < (n-1)+\prod_{j=1}^n d_j
    \end{equation}
    holds, then $\mathcal{H}$ is not universally locally encodable.
\end{prop}
Note here that the criterion of Pauwels and Gühne on the dimensions of $\mathcal{H}_j$ in Eq.~\eqref{ineq 2} appears to be slightly weaker than the negation of the naive bound from Eq.~\eqref{eq:dim_ineq}, but is in fact equivalent (see Appendix Remark~\ref{vieta}).
Additionally, the one-uniform states are the (generalized) locally maximally entangled states from Ref.~\cite{Bryan} (not to be confused with the locally maximally entanglable states, that appeared in Proposition~\ref{prop:group_summary}(b)). These are states with ${\Tr_{\{1,...,n\}\backslash\{j\}}(\ket{\psi}\bra{\psi}) = \mathbbm{1}/d_j}$ for all $j \in \{1,...,n\}$. It is shown in Ref.~\cite{Bryan,bryan2018existence} in Theorem~1.2 that there is a continuum of such states (up to $\LU$ equivalence) if and only if $\sum\limits_{j=1}^n d_j^2 < (n+1) + \prod\limits_{j=1}^nd_j$. 
This turns out to be equivalent to the negation of inequality~\eqref{eq:dim_ineq} as well by a similar argument.

It remains to show that one-uniform, regular states always exist if Eq.~\eqref{eq:dim_ineq} is violated. To that end, the criterion of regularity for a one-uniform state can be rephrased in terms of its stabilizer group.
\begin{prop}[Appendix Proposition~\ref{stab}]
    Let $\ket{\psi} \in \mathcal{H}$ be a one-uniform state. Then it is regular if and only if its stabilizer Lie-group
    \begin{equation*}
        S_{\ket{\psi}}=\left\{ M \in \bigtimes_{j=1}^n \SL(d_j,\C) : M\ket{\psi} = \ket{\psi} \right\}
    \end{equation*}
    is discrete (it fulfills $\dim(S_{\ket{\psi}}) = 0$).
\end{prop}
Under the assumed violation of inequality \eqref{eq:dim_ineq}, both the set of states with discrete stabilizer and those related to one-uniform states by SLOCC are in some sense large enough, that a nontrivial overlap can always be guaranteed. We show this in Appendix Lemma~\ref{density}. This culminates in:

\begin{theorem}[Appendix Theorem~\ref{th:impossible}]\label{th:impossible_main}
    Let $\mathcal H = \bigotimes_{k=1}^n \mathbb{C}^{d_k}$. If 
    \begin{align}
        \sum\limits_{j=1}^n d_j^2 < \prod\limits_{j=1}^n d_j+n,
    \end{align}
    then $\mathcal H$ is not universally locally encodable.
\end{theorem}
We display some choices of local dimensions and the implied results on universal local encodability in Tab.~\ref{tab:constructions2}.

\section{Application}
\label{sec:application}

While the problem of universal local encodability is of intrinsic interest for our understanding of the geometry of multipartite quantum systems, it also has an operational interpretation for the task of covariant quantum estimation, a standard setting for estimating unknown transformations
and entangled states~\cite{Chiribella2004Covariant,Chiribella2005Group,Sacchi2006Tradeoff}.

To that end, consider a known probe state $\rho_0$ on
$\mathcal H=\bigotimes_{k=1}^{N}\mathbb C^{d_k}$, with
$D=\prod_k d_k$.
The probe undergoes an unknown local unitary rotation
$V=\bigotimes_k U_k$, giving $\rho_V=V\rho_0V^\dagger$.
We receive one copy of $\rho_V$ and report an estimate $\widehat V$.
The local unitaries are drawn independently from the normalized Haar
measure on $\LU$. In order to quantify the success probability, consider any covariant loss function $\ell$, i.e., we can write
$\ell(\widehat V,V)=c(\widehat V^{\dagger}V)$.

Standard choices for $c$ include, for pure probe states $\rho_0 = \ketbra{\psi}{\psi}$, the mean infidelity~\cite{Massar1995Estimation, Bruss1999Estimation}
\begin{align}
    c_F(V)=1-|\braket{\psi|V}{\psi}|^2,
\end{align}
and the probability of exceeding a fixed trace-distance
tolerance when estimating $\rho_V$, given by \cite{Keyl2006LargeDeviations}
\begin{align}\label{eq:ceps}
    c_\varepsilon(V) = \begin{cases}1 & \frac12\Vert \rho_0 - V\rho_0 V^\dagger \Vert_1 > \varepsilon, \\ 0 & \text{otherwise}\end{cases},
\end{align}
where $\Vert \cdot \Vert_1$ denotes the trace norm. 

We are interested in the POVM measurement $M$ with outcomes $\widehat V$ that minimizes the average loss
\begin{align}\label{eq:R(M)}
    R(M,\rho_0)&=\mathbb E[\ell(\widehat V,V)] \nonumber \\
    &= \int_{\LU} \text{d}V \int_{\LU} \text{d}\widehat V\,  \ell(\widehat V,V) \Tr(\rho_V M_{\widehat V}).
\end{align}

We show that universal local encodability guarantees an optimal
projective measurement (in contrast to general POVMs) for such estimation tasks:
\begin{theorem}\label{thm:LUdisc}
    Let $\mathcal H=\bigotimes_{k=1}^{N}\mathbb C^{d_k}$ be finite dimensional and universally locally encodable. Consider one copy of $\rho_V=V\rho_0V^\dagger$, where $\rho_0$ is a known probe state and $V \in \LU$ is a Haar-distributed local unitary. For every bounded invariant loss function 
    \begin{align*}
        \ell(\widehat V,V)=c(\widehat V^{\dagger}V),
    \end{align*}
    the minimum average loss $\min_M R(M,\rho_0)$ over all POVMs $M$ is attained by a single rank-one projective measurement on $\mathcal H$.
\end{theorem}
\begin{proof}
Covariance allows us to choose an optimal POVM of the form
$M_{\widehat V} = \widehat V\Xi\widehat V^\dagger$ for some $\Xi\geq0$ and $\Tr\Xi=D$, such that $M_{\widehat V} \geq 0$ and $\int M_{\widehat V}\, \text{d}\widehat V = \id$ (see~\cite{Chiribella2005Group}).

Defining the cost operator
\begin{equation}
 C(\rho_0)=\int_{\LU} c(U)\,U\rho_0U^\dagger\,\text{d}U,
 \label{eq:app-cost-operator}
\end{equation}
we can write $R(M,\rho_0)=\Tr[\Xi C(\rho_0)]$. 
Thus, the minimum over all POVMs is given by
\begin{align}
 R_{\rm POVM}(\rho_0)&:=\min_{M\in\text{ POVM}} R(M,\rho_0) \nonumber \\
 &= \min_{\Xi \geq 0,\Tr(\Xi)=D} \Tr[\Xi C(\rho_0)] \nonumber \\
 &=D c_{\min}, 
 \label{eq:app-optimal-risk}
\end{align}
with $\Xi_{\mathrm{opt}}=D\ketbra{\eta}{\eta}$, and $c_{\min}$ is the smallest eigenvalue of $C$ and
$\ket{\eta}$ is a corresponding normalized eigenvector.

Now consider a projective measurement $P$ in an orthonormal basis
$\{\ket{e_a}\}_{a=1}^{D}$, which reports  the estimated local unitary $\widehat V_a$  upon measuring outcome $a$, i.e., $P_{\widehat V} = \sum_{a=1}^D \delta(\widehat V_a^\dagger\widehat V) \ketbra{e_a}{e_a}$. Then Eq.~\eqref{eq:R(M)} yields
\begin{equation}
 R(P)
 = \sum_{a=1}^{D}
   \braket{e_a | \widehat V_aC \widehat V_a^\dagger}{e_a}\geq D c_{\min}.
 \label{eq:app-projective-risk}
\end{equation}
If $\ket{\eta}$ is locally encodable with encoders $\widehat V_a$, we can choose $\ket{e_a} =\widehat V_a \ket{\eta}$.
Every term then equals $c_{\min}$, so this projective measurement
attains the optimum over all POVMs.
Universal local encodability for a given Hilbert space therefore guarantee an
optimal projective measurement for every probe state and invariant loss function. Note that the encoded $\ket{\eta}$ depends on the probe state and the chosen loss function and generally does not coincide with the probe state.
\end{proof}

If we choose as cost function the fixed trace-distance tolerance $c_\varepsilon$ from Eq.~\eqref{eq:ceps}, also the converse is true, provided $\varepsilon>0$ is chosen sufficiently small. To that end, define
\begin{align*}
    R_{\rm POVM}^\varepsilon(\rho_0)&:=\!\!\!\!\!\min_{M\in \text{POVM}}   R^\varepsilon(M,\rho_0) \\
    &= \!\!\!\!\!\min_{M\in \text{POVM}} \int_{\LU} \!\!\!\!\text{d}V \!\!\int_{\LU} \!\!\!\!\text{d}\widehat V\,  c_\varepsilon(\widehat V^{\dagger}V) \Tr(\rho_V M_{\widehat V}),
\end{align*}
and likewise $R_{\rm PVM}^\varepsilon$ where we minimize over projective measurements instead. 
In Appendix~\ref{ap:POVM}, we show:
\begin{lemma}\label{lem:LUdisceps}
    Let $\mathcal H=\bigotimes_{k=1}^{N}\mathbb C^{d_k}$ be finite dimensional and $\ket{\psi}\in \mathcal{H}$ not locally encodable. Then there exists an $\varepsilon > 0$ such that
    \begin{align}
        R_{\rm POVM}^\varepsilon(\ketbra{\psi}{\psi}) <         R_{\rm PVM}^\varepsilon(\ketbra{\psi}{\psi}).
    \end{align}
\end{lemma}
Note that relaxing the optimization over projective measurements to include projectively simulable measurements cannot close the gap, as the optimization is convex~\cite{Oszmaniec2017Simulation}.

Lemma~\ref{lem:LUdisceps}, together with Theorem~\ref{thm:LUdisc}, implies:
\begin{corollary}
    Let $\mathcal H=\bigotimes_{k=1}^{N}\mathbb C^{d_k}$ be finite dimensional. Then $\mathcal{H}$ is universally locally encodable, iff for all $\ket{\psi} \in \mathcal{H}$ and all $\varepsilon > 0$, $R_{\rm POVM}^\varepsilon(\ketbra{\psi}{\psi}) = R_{\rm PVM}^\varepsilon(\ketbra{\psi}{\psi})$. 
\end{corollary}

\section{Discussion}
In this paper we provided a large collection of new results related to the question of universal local encodability, as summarized in Tables~\ref{tab:constructions} and~\ref{tab:constructions2}. We deepened the understanding of known results by finding a similarity in the previously explored constructions of $2\times2,4\times 4$ and $8 \times 8$ from Ref.~\cite{PRT}, and showed that they cannot be canonically generalized (Proposition~\ref{sym limit}). Still we were able to lift these constructions to $d_A \times d_B$ for any $d_A \geq d_B$ and $d_B \in \{2,4,8\}$ (Theorem~\ref{th:dx2_ext}). We also found lifting theorems for other arbitrary constructions (Lemma~\ref{lem:enc_poss_multi}).

We introduced the infinite dimensional variant of the universal local encoding problem (Section~\ref{subsec:inf_dim}), and resolve it fully in the bipartite setting using Hurwitz-Radon matrices and a lifting result (Proposition~\ref{prop:Radon} and Theorem~\ref{th:oo_x_oo_main}). 
It would be interesting to generalize our findings to the multipartite setting with at least one infinite dimensional part, since no easy heuristic based on dimension counting exists in these cases.

In order to show universal local encodability in a structured way, we introduced a group construction for $n$-party systems that generates a large set of maximally coherent locally encodable states (Propositions~\ref{prop:group} and~\ref{prop:group_summary}). We showed that for coprime dimensions in bipartite systems, the only groups satisfying this structure are cyclic (i.e. fairly trivial), yet this is surprisingly sufficient for multiple dimensional pairs including $4\times 3$ (see Appendix Section~\ref{4x3 app}). It would be interesting to know, for what other dimensional tuples this sufficiency of the cyclic group holds true. Any nontrivial structure requires $\gcd(d_A,d_B) > 1$. 

For $d_A = d_B = 3$ we were able to proof universal local encodability using this method (explicitly outlined in Appendix Section~\ref{3x3}). It might be possible to show that this approach suffices to prove universal local encodability in any such case, or at least for $d \times d$, since most (but crucially not all) steps of the proof are already presented in a generalized form. This is formally stated as Conjecture~\ref{conj:group}.

There also exists a nontrivial group for three-qubit systems, given by $G = \langle Z^{\otimes 3},X\otimes \id \otimes X, \id \otimes X \otimes X \rangle$, whose eigenstates are a $\LU$ rotation of the $\ket{\GHZ}$ basis. The corresponding maximally coherent states might be an $\LU$-universal set, which would resolve the remaining conjecture from Ref.~\cite{PRT} that all three qubit states are locally encodable. More generally, it would be interesting to characterize the existence and structure of these groups and the corresponding maximally coherent states, especially with regard to universal local encodability.

In Section~\ref{sec:nogo}, we generalize the no-go theorem of Pauwels and Gühne in Ref.~\cite{Guehne} to all local dimension tuples where non-ULE was already suggested based on the violation of a heuristic dimensional inequality (see Section~\ref{subsec:heu} and Theorem~\ref{th:impossible_main}). The converse of this claim, that fulfillment of the inequality yields universal local encodability, cannot be established using this method and remains an open problem.

Finally, we gave operational meaning to the feature of ULE by showing that it is equivalent to the existence of an optimal projective measurement for a class of invariant estimation task. Finding further implications of ULE, for instance for the geometry of projectively simulable POVMs, appears to be a fruitful future direction of research.

\begin{acknowledgments}
We thank Raphael Brinster, Dagmar Bruß, Otfried Gühne and Tulja Varun Kondra for fruitful discussions. J.E.~acknowledges
financial support by Deutsche Forschungsgemeinschaft (DFG, German Research
Foundation) under Germany's Excellence Strategy -- Cluster of Excellence
Matter and Light for Quantum Computing (ML4Q) EXC 2004/2 -- 390534769.

Showing equivalence of universal local encodability with optimality of projective measurements in discrimination tasks, connecting regularity of a state to its stabilizer and some parts of the constructions, in particular finding phases on the boundary $\Gamma_1$ for the two-qutrit construction, finding the counterexample for Proposition~\ref{prop:group_summary}(c) and the limits of Proposition~\ref{th:dx2_ext}, parts of the infinite-dimensional Hurwitz-Radon construction, as well as some technical steps in the proofs have been assisted by use of ChatGPT (OpenAI, GPT-5.6 Sol and GPT-6.1 Sol; accessed July-November 2026).
\end{acknowledgments}
\bibliographystyle{apsrev4-1}
\bibliography{refs}
\appendix
\onecolumngrid
\parindent 0pt
\section{The Hurwitz-Radon Construction}\label{Ap HR}
\subsection{Finite Dimensions}
The Hurwitz-Radon Construction only works on states of a certain form. It must first be checked, that they are $\LU$-universal.
    \begin{lemma}\label{Radon uni lemma}
        Let $d_A \geq d_B \in \N_{\geq 2}$ and $U \in \C^{d_A\times d_B}$ be an isometry. Then the set
        \begin{equation*}
            \left\{ \ket{\psi} = \frac{1}{\sqrt{d_B}}\sum_{j=0}^{d_B-1} U\ket{v_j} \otimes \ket{j} \in \C^{d_A} \otimes \C^{d_B} : \ket{v_j} \in \R^{d_B}, |\ket{v_j}|=1 \forall j \in \{0,...,d_B-1\} \right\}
        \end{equation*}
        is $\LU$-universal.
    \end{lemma}
    \begin{proof}
        It is sufficient to construct a set of normed vectors $(\ket{v_j})_j$ for every set of Schmidt values. Equivalently $\varrho_B = \Tr_A(\ket{\psi}\bra{\psi})$ has to have theirs squares as eigenvalue. Now fix such a set $(s_0^2,...,s_{d_B-1}^2)$. By the symmetric Schur-Horn theorem (the first theorem in Ref.~\cite{SchurHorn}), there is a real positive symmetric matrix $C$ with $\diag(C)= \left(\frac{1}{d_B},...,\frac{1}{d_B}\right)$ and these eigenvalues. Denote the $j$-th column of $\sqrt{C}$ by $v_j/\sqrt{d_B} \in \R^{d_B}$. Then $C_{jk} = \frac{\braket{v_j}{v_k}}{d_B}$ and thus $|\ket{v_j}|^2 = 1$. On the other hand $
            \varrho_B = \frac{1}{d_B}\sum\limits_{j,k=0}^{d_B-1} \braket{v_j}{U^\dagger U v_k} \ket{j} \bra{k} = C$.
    \end{proof}
    \begin{prop}
        Fix $d_B \in \N_{\geq 2}$ and choose $d_A \in \N_{\geq d_B}$ such that $\varrho(d_A) \geq d_B$, where $\varrho(\cdot)$ denotes the Radon number. Then universal local encodability is possible for $d_A \times d_B$. Furthermore, for all $d_B \in \N_{\geq 2}$, local encodability is possible for $\infty \times d_B$.
    \end{prop}
    \begin{proof}
        Let $\{A_j \in \R^{d_A \times d_A}: j=0,...,d_B-1\}$ denote a $d_B \leq \varrho(d_A)$ sized subset of the Hurwitz-Radon matrices. 
        Furthermore let $E_{\ket{s}} = \begin{bmatrix}A_0 \ket{s} & \cdots & A_{d_B-1} \ket{s} \end{bmatrix} \in \R^{d_A \times d_B}$. Since $A_j^\dagger A_k$ is an odd matrix for $j\neq k$ and $\mathbbm{1}$ else, $E_{\ket{s}}$ is an isometry. Furthermore for all $s\neq s'$ the matrix $E_{\ket{s'}}^\dagger E_{\ket{s}}$ is skew-symmetric. There exists an orthogonal $U_{\ket{s}} \in \R^{d_A \times d_A}$, such that $E_{\ket{s}} = U_{\ket{s}} E_{\ket{0}}$, obtained by first using Gram-Schmidt to extend the isometries to unitaries $\tilde{E}_{\ket{s}} \in \R^{d_A \times d_A}$, and then setting $U_{\ket{s}} = \tilde{E}_{\ket{s}} \tilde{E}_{\ket{0}}^\dagger$.
        By Lemma~\ref{Radon uni lemma} it is sufficient to give an encoder for any $\ket{\psi} = \frac{1}{\sqrt{d_B}} \sum\limits_{j=0}^{d_B-1} E_{\ket{0}} \ket{v_j} \otimes \ket{j}$.
        It is given by the $d_Ad_B$ sized set $\{U_{\ket{j}} \otimes Z^k : j \in \{0,...,d_A\}, k \in \{0,...,d_B\} \}$, since
        \begin{equation*}
        \begin{split}
            \braket{\psi}{(U_{\ket{\alpha}}^\dagger U_{\ket{\beta}} \otimes Z^{\gamma-\delta})\psi} &= \frac{1}{d_B} \sum_{j,k=0}^{d_B-1} \braket{v_j}{E_{\ket{0}}^\dagger U_{\ket{\alpha}}^\dagger U_{\ket{\beta}} E_{\ket{0}}v_k} \braket{j}{k}= \frac{1}{d_B} \sum_{j,k=0}^{d_B-1} \braket{v_j}{E_{\ket{\alpha}}^\dagger E_{\ket{\beta}} v_k} \omega_{d_B}^{(\gamma-\delta)j} \delta_{jk} \\
            &=\frac{1}{d_B} \sum_{j=0}^{d_B-1} \omega_{d_B}^{(\gamma-\delta)j} \underbrace{\braket{v_j}{E_{\ket{\alpha}}^\dagger E_{\ket{\beta}} v_j}}_{=\delta_{\alpha,\beta}} =\delta_{\alpha,\beta} \delta_{\gamma,\delta}
        \end{split}
        \end{equation*}
        where the skew-symmetry of $E^\dagger_{\ket{\alpha}} E_{\ket{\beta}}$ and the realness of $v_j$ was used.
        The $d_A = \infty$ case is an application of Lemma~\ref{lem:enc_poss_multi}.
    \end{proof}
    Note that $\varrho(2^k) \geq 2k$ for all $k$, so that $\varrho(d_A=2^{\lceil d_B/2 \rceil}) \geq d_B$. The optimal choice of $d_A$ is listed in Table~\ref{tab:radon}.
\begin{table}[t]
    \begin{tabular}{l||c|c|c|c|c|c|c|c|c|c|}
      $d_B$ &  $2$ & $3$  $4$ &$5$ $6$ $7$ $8$ & $9$ & $10$ & $11$  $12$ & $13$ $14$ $15$ $16$ & $17$ & $18$ & $19$ $20$\\
      \hline
      $d_A$ & $2$ & $4$ & $8$  & $16$ & $32$ & $64$  & $128$& $256$ & $512$ & $ 1024$\\
    \end{tabular}
    \caption{List of optimal $d_A$ values using the Hurwitz-Radon construction for each $d_B$. This corresponds to the OEIS list~\cite{oeis}.}
    \label{tab:radon}
\end{table}
\subsection{Infinite Dimensions}\label{HR oo}
\begin{lemma}\label{Schmidt oo}
    Every normalized $\ket{\psi} \in \ell^2 \otimes \ell^2$ is $\LU$ equivalent to a state of the following form (here $N,M \in \N_0$, $(s_j)_j$ positive with $\sum_j s_j^2 = 1$)
    \begin{equation*}
            (1) \; \sum_{j=0}^N s_j \ket{j,j}, \quad (2) \; \sum_{j=0}^\infty s_j \ket{j+M,j+N}, \quad (3) \; \sum_{j=0}^\infty s_j \ket{j+M,2j}, \quad
            (4) \; \sum_{j=0}^\infty s_j \ket{2j,j+N}, \quad(5) \; \sum_{j=0}^\infty s_j \ket{2j,2j}
    \end{equation*}
\end{lemma}
\begin{proof}
    The Schmidt-decomposition $\mathcal{H}_1 \otimes \mathcal{H}_2$ for two separable Hilbert spaces yields two orthonormal sets of vectors $(\ket{e_j})_j \subset \Hl_1$ and $(\ket{f_j})_j \subset \Hl_2$ (of each at most countable size) and strictly positive numbers $(s_j)_j$, such that
    $\ket{\psi} = \sum_j s_j \ket{e_j} \otimes \ket{f_j}$. The five options correspond to different dimensions of $E= \overline{\vspan\{\ket{e_j}\}}, F=\overline{\vspan\{\ket{f_j}\}}$ as well as $E^\perp$ and $F^\perp$.
\end{proof}
In particular one can always consider states of the form $\ket{\psi}=\sum\limits_{j=0}^\infty s_j \ket{a_j',b_j'}$, where $\{a_j'\},\{b_j'\}$ are both subsets of the canonical basis (with $a_j'$ and $b_j'$ strictly monotonically increasing) and $s_j \geq 0$.
\begin{rem}
    Consider the set of binary strings with finitely many nonzero terms 
    \begin{equation*}
        c_{00}(\N_0,\mathbb{F}_2) = \{ x=(x_n)_{n \in \N_0} \in \mathbb{F}_2^{\N_0}: x_n=0 \mathrm{ \;for\; all \;but\; finitely \;many\; } n\}
    \end{equation*}
    equipped with the element-wise addition $\oplus$. Note that $c_{00}(\N_0,\mathbb{F}_2)$ is a vector space over $\mathbb{F}_2$. There is a bijection between $c_{00}(\N_0,\mathbb{F}_2)$ and $\N_0$ given by 
    \begin{equation*}
        \Phi : c_{00}(\N_0,\mathbb{F}_2) \to \N_0, \qquad x \mapsto \sum\limits_{j=0}^\infty 2^j x_j
    \end{equation*}
    This induces an ordering on $c_{00}(\N_0,\mathbb{F}_2)$ via $x<y \iff \Phi(x)<\Phi(y)$.
    Conversely $\Phi$ induces an operation $\oplus$ on $\N_0$ via $a\oplus b = \Phi(\Phi^{-1}(a) \oplus \Phi^{-1}(b))$. For example $17 \oplus 13 = 28$.
\end{rem}
\begin{lemma}\label{lin indep}
    Let $(a_j)_{j \in \N_0}, (b_j)_{j \in \N_0}$ be strictly monotonically increasing sequences in $c_{00}(\N_0,\mathbb{F}_2)$. Then there is a permutation $\sigma : \N_0\to \N_0$ such that the set $\{ (a_j,b_{\sigma(j)}) : j \in \N_0\} \subseteq c_{00}(\N_0,\mathbb{F}_2)^{\oplus 2}$ consist of linearly independent vectors.
\end{lemma}
\begin{proof}
    The permutation is inductively defined. For this let $\mathcal{B}_N$ be the sets of pairs $(a_j,b_{\sigma(j)})$, for which $\sigma(j)$ has been assigned in steps $1$ to $N$. Let $V_N = \vspan\{\mathcal{B}_N\}$. Note that due to linear independence $\# \mathcal{B}_N = N$ and $\# V_N = 2^{N}$.\\
    \underline{Step 1:} Letting $\sigma(0)=0$ can be a problem if $a_0=b_0=0$, since then $(a_0,b_0)$ is the zero vector which is linearly dependent on everything. Since $(a_j)_j$ and $(b_j)_j$ are strictly monotonically increasing, setting $\sigma(0)=1$ fixes this problem. This defines $\mathcal{B}_1$ and $V_1$. Since $V_0=\mathcal{B}_0 \cup \{(0,0)\}$ and $(a_1,b_0) \notin \mathcal{B}_0$ one can further set $\sigma(1)=0$ as Step $2$.
    \underline{Induction:} Now use the following strategy to get from even $N$ to $N+2$. \\
    Step $N+1$: Let $\pi_k:c_{00}(\N_0,\mathbb{F}_2)^{\oplus 2} \to c_{00}(\N_0,\mathbb{F}_2)$ ($k\in \{1,2\}$) be the projection on the $k$-th coordinate.
    Chose the first $j$ where $\sigma(j)$ is still unassigned, and assign it the smallest $k$, such that $b_k \notin \pi_2(V_N)$ (which exists since $V_N$ is finite and $(b_j)_{j \in \N_0}$ is an infinite strictly increasing sequences)\\
    Step $N+2$: Now choose the smallest $k$ where $\sigma^{-1}(k)$ is still unassigned and assign it the smallest $j$ such that $a_j \notin \pi_1(V_{N+1})$.\\
    This back and forth guarantees that after $2(N+1)$ steps $N$ is both in the domain and the range of $\sigma$, and no input or output is assigned twice.
\end{proof}
\begin{corollary}\label{f_j col}
    Let $(a_j)_{j \in \N_0}, (b_j)_{j \in \N_0}$ be strictly monotonically increasing sequences in $c_{00}(\N_0,\mathbb{F}_2)$ and choose $\sigma$ according to Lemma~\ref{lin indep}.
    For every $j \in \N_0$ there exists a linear function  $f_j : c_{00}(\N_0,\mathbb{F}_2)^{\oplus 2} \to \mathbb{F}_2$, such that $f_j((a_k,b_{\sigma(k)})) = \chi_{\{j > k\}}$.
\end{corollary}
\begin{proof}
    Since the vectors $\{(a_k,b_{\sigma(k)}) : k \in \N_0\}$ are independent, a linear function can obtain arbitrary values on them. To define a function on the total domain extend $\bigcup\limits_{N \in \N_0} \mathcal{B}_N$ to a basis, and assign $f_j$ to be zero on all additional basis elements.
\end{proof}
\begin{lemma}\label{oo Hurwitz}
    Let $(a_j)_{j \in \N_0}, (b_j)_{j \in \N_0}$ be strictly monotonically increasing sequences in $c_{00}(\N_0,\mathbb{F}_2)$ and choose $\sigma$ according to Lemma~\ref{lin indep}.
    For $j \in \N_0$ define $f_j$ via Corollary~\ref{f_j col} and let $A_j \in L(\ell^2)$ be defined via 
    \begin{equation*}
        A_j \ket{k,m} = (-1)^{f_j(\Phi^{-1}(k),\Phi^{-1}(m))} \ket{k\oplus \Phi(a_j), m \oplus \Phi(b_{\sigma(j)})}.
    \end{equation*}
    Then set of operators $\{A_j : j \in \N_0\}$ satisfies the Hurwitz-Radon-like condition 
        $A_j^\dagger A_k+A_k^\dagger A_j = 2 \delta_{j,k} \mathbbm{1}$.
\end{lemma}
\begin{proof}
    The adjoint of $A_j$ is easily seen to be given by
    \begin{equation*}
        A_j^\dagger \ket{k,m}= (-1)^{f_j(\Phi^{-1}(k)\oplus a_j,\Phi^{-1}(m)\oplus b_{\sigma(j)}} \ket{k\oplus \Phi(a_j), m \oplus \Phi(b_{\sigma(j)})}.
    \end{equation*}
    Now let $\ket{p',q'} = (-1)^{f_j(\Phi^{-1}(p),\Phi^{-1}(q))+f_k(\Phi^{-1}(p),\Phi^{-1}(q))}\ket{p\oplus \Phi(a_j)\oplus \Phi(a_k),q\oplus\Phi(b_{\sigma(j)})\oplus \Phi(b_{\sigma(k)})}$, then
    \begin{equation*}
    \begin{split}
        (A_j^\dagger A_k + A_k^\dagger A_j)\ket{p,q} =\left((-1)^{f_j(a_k,b_{\sigma(k)})+f_j(a_j,b_{\sigma(j)})}+(-1)^{f_k(a_k,b_{\sigma(k)})+f_k(a_j,b_{\sigma(j)})}\right)\ket{p',q'}
    \end{split}
    \end{equation*}
    using linearity of $f_j$. If $j<k$ or $j>k$ then the factor in front of $\ket{p',q'}$ is given by $0$. For $j=k$ the prefactor is $2$ and $\ket{p',q'} = \ket{p,q}$.
\end{proof}
\begin{theorem}\label{oo x oo enc}
    Every $\ket{\psi} \in \ell^2 \otimes \ell^2$ is locally encodable.
\end{theorem}
\begin{proof}
    Write an $\ket{\psi}$ in an $\LU$ equivalent form from Lemma~\ref{Schmidt oo}. Using the canonical link between $c_{00}(\N_0,\mathbb{F}_2)$ and $\N_0$ it can be written as
        $\ket{\psi'} =\sum\limits_{j=0}^\infty s_j \ket{\Phi(a_j),\Phi(b_j)}$
    for strictly monotonically increasing sequences in $c_{00}(\N_0,\mathbb{F}_2)$. Let $(A_j)_{j \in \N_0}$ be the operators from Lemma~\ref{oo Hurwitz} for these sequences $(a_j)_{j \in \N_0},(b_j)_{j \in \N_0}$. Choose $\sigma$ according to Lemma~\ref{lin indep} and define the unitary operator $U_\sigma \in L(\ell^2)$ via $U_\sigma\ket{\Phi(b_j)} = \ket{\Phi(b_{\sigma(j)})}$, and as $\id$ on $\overline{\vspan\{\ket{\Phi(b_j)} : j \in \N_0\}}^\perp$. Let $\ket{\psi''} = (\id \otimes U_\sigma) \ket{\psi'}$. It is sufficient to find encoders for $\ket{\psi''}$. Let
    \begin{equation*}
        X_p \ket{j} = \ket{j\oplus p}, \quad D_{p,q}\ket{\Phi(a_j)} = (-1)^{f_j(\Phi^{-1}(p),\Phi^{-1}(q))} \ket{\Phi(a_j)}
    \end{equation*}
    for $f_j$ defined via Corollary~\ref{f_j col} and extend $D_{p,q}$ by $\id$ on $\overline{\vspan\{\ket{\Phi(a_j)} : j \in \N_0\}}^\perp$.
    Let
    \begin{equation}
        \ket{\psi_{p,q}} = (X_pD_{p,q} \otimes X_q) \ket{\psi''}= \sum_{j=0}^\infty s_j (-1)^{f_j(\Phi^{-1}(p),\Phi^{-1}(q))} \ket{\Phi(a_j) \oplus p,\Phi(b_{\sigma(j)})\oplus q}=\sum_{j=0}^\infty s_j A_j \ket{p,q}
    \end{equation}
    The Operator $\tilde{U}=\lim\limits_{N\to \infty} \sum\limits_{j=0}^N s_j A_j$ is well defined with operator norm convergence, since for any $v \in \ell^2$ with $||v||=1$
    \begin{equation}\label{A converges}
        ||\sum\limits_{j=N}^M s_j A_j v||_{\ell^2}^2 = \!\!\sum\limits_{j,k=N}^M s_j s_{k} \braket{v}{A_j^\dagger A_k v}   
        = \!\sum_{j=N}^M s_j^2 ||v||^2 + \!\sum_{\substack{j=N\\k >j}}^M s_js_j' \braket{v}{(A_j^\dagger A_k +A_k^\dagger A_j) v} =\!\! \sum_{j=N}^M s_j^2
    \end{equation}
    is bound independent of $v$ and a Cauchy-sequence, by convergence of $\sum_j s_j^2$ to 1. The collapse of the double sum is due to the Hurwitz-Radon like condition. Note that $\ket{\psi_{p,q}} = \tilde{U}\ket{p,q}$. It thus remains to show, that $\tilde{U}$ is unitary. $\tilde{U}$ is an isometry by Equation (\ref{A converges}) (take $N=0$ and $M \to \infty$). Furthermore $\tilde{V} \coloneqq \lim\limits_{N\to \infty} \sum\limits_{j=0}^N s_j A_j^\dagger$ is also easily seen to be well defined and an isometry. Additionally, $\tilde{U}$ and $\tilde{V}$ are bounded and thus continuous, such that 
    \begin{equation*}
        \tilde{U}\tilde{V} \ket{v} = \lim_{N\to \infty} \sum_{j,k=0}^Ns_js_k A_j A_k^\dagger \ket{v} = \lim_{N\to \infty} \sum_{j=0}^N s_j^2 \ket{v} = \ket{v} = \tilde{V}\tilde{U}\ket{v}
    \end{equation*}
    and thus $\tilde{V}\tilde{U}^{-1}$. Since $\tilde{U}$ is an invertible isometry it is a unitary operator. Now $\ket{\psi} \stackrel{\LU}{\sim} \ket{\psi''} \stackrel{\LU}{\sim} \ket{\psi_{p,q}}$ and the set of all $\ket{\psi_{p,q}}$ is a unitary rotation of the canonical basis and thus a complete orthonormal basis itself.
\end{proof}

\section{Group Approach - Abstract Constructions}\label{Ap group}
\subsection{General Constructions}
\begin{lemma}\label{joint eig}
    Let $G$ be a group satisfying the properties of Proposition~\ref{prop:group}. Then the joint eigenspaces of all group elements are one dimensional.
\end{lemma}
\begin{proof}
    Consider a decomposition of $\mathcal{H} = \bigoplus\limits_\lambda E_\lambda$, such that each element of $g\in G$ acts as $\lambda^{(g)} \mathbbm{1}$ on $E_\lambda$.
    The goal is to show $m_\lambda := \dim(E_\lambda)\stackrel{!}{=}1$ for all $\lambda$. Since $g\mapsto \lambda^{(g)}$ is a character, the Schur orthogonality relations imply 
    $\frac{1}{\#G}\sum\limits_{g\in G} \frac{\mu^{(g)}}{\lambda^{(g)}} = \delta_{\lambda,\mu}$.
    This now yields (using $\sum_\mu m_\mu \mu^{(g)} =\Tr(g) = \#G \delta_{g,\mathbbm{1}}$ and $\lambda^{(\mathbbm{1})} = 1$):
    \begin{equation*}
        1  = \frac{1}{\#G}\sum_{g\in G} \frac{\Tr(g)}{\lambda^{(g)}} = \frac{1}{\#G}\sum_{g\in G} \sum_{\mu} \frac{m_\mu\mu^{(g)}}{\lambda^{(g)}} = \sum_\mu m_\mu \delta_{\mu,\lambda} =m_{\lambda}.\qedhere
    \end{equation*}
\end{proof}
\begin{examp}\label{prod ex}
    The simplest kind of abelian groups of order $D$ is the cyclic group $\mathbb{Z}_{D}$ generated by one Element $g$ with $\ord(g)=D$. Indeed this group can be realized using traceless local unitary matrices. Let $\omega_d = e^{2\pi i/d}$ be the $d$-th root of unity, and set $d_k' = \prod\limits_{\ell=1}^k d_\ell$. Then 
    \begin{equation}\label{prod const}
        g = \bigotimes_{k=1}^n \begin{bmatrix} \omega_{d_k'}^0 &  & &\\ & \omega_{d_k'}^1& & \\ & & \ddots  & \\ & & & \omega_{d_k'}^{d_k-1}\end{bmatrix}.
    \end{equation}
    is of order $D$ and $G=\brakett{g}$ satisfies Proposition~\ref{prop:group}. Its eigenbasis consists of the product states. If $\left\{\ket{\vec{\xi}}\right\}_{\vec{\xi}}$ is a product basis of $\Hl = \bigotimes\limits_{k=1}^n \Hl_j$, then by Proposition~\ref{prop:group} any state $\LU$-equivalent to one of the form
    \begin{equation}\label{LME like}
        \ket{\psi(\theta)} = \frac{1}{\sqrt{d_{\Hl}}} \sum_{\xi} e^{i\theta_\xi} \ket{\xi}
    \end{equation}
    for some $(\theta_\xi)_{\xi} \subset \R$ is locally encodable.
\end{examp}
\begin{rem}\label{LME rem}
In the case of $d_k=2$ for all $k=1,...,n$, states of the form (\ref{LME like}) are the locally maximally entanglable (LME) states (see Theorem 2 in Ref.~\cite{Kraus}). It is well known, that they are locally encodable (compare Lemma 1 in \cite{Kraus}), a result that now easily follows from the group approach. For $d_j \equiv 2, n \geq 3$, this group cannot be sufficient, as the state $\ket{W_n} = \frac{1}{\sqrt{n}} \sum\limits_{j=0}^{n-1} \ket{\underbrace{0...0}_{j}1\underbrace{0...0}_{n-j-1}}$
is not $\LU$-equivalent to a state of the form (\ref{LME like}) according to Proposition 3 in \cite{Wstate}. This is not surprising for $n \geq 4$ by the degree of freedom count, but for $n=3$ this motivates to look beyond cyclic groups. Even in the bipartite case, their maximally coherent states are not always $\LU$-universal. Consider for example the state $\ket{\varphi} = \frac{\ket{00}+\ket{11}}{\sqrt{2}}$ which is not $\LU$-equivalent to a state of the form (\ref{LME like}) for $n=2, d_A\geq 3$ odd and $d_B=3$.
\end{rem}
\begin{proof}
    The nonzero Schmidt coefficients of $\ket{\varphi}$ are given by $\left(\frac{1}{\sqrt{2}},\frac{1}{\sqrt{2}}\right)$. Assume there is a $\theta \in \R^{d_A\times 3}$, such that $\ket{\psi(\theta)}$ is $\LU$-equivalent to $\ket{\varphi}$.
    Let $(A_{jk})_{j,k} = \frac{1}{3} e^{i \theta_{jk}}$ and $B = A^\dagger A$.
    By the $\LU$-equivalence $B \in \C^{3\times 3}$ must have eigenvalues $\left(\frac{1}{2},\frac{1}{2},0\right)$, and thus $B = \frac{\mathbbm{1}}{2} - \frac{1}{2} \ket{v}\bra{v}$ for a normed vector $v$. This yields $\frac{1}{3} = (A^\dagger A)_{kk} = B_{kk} = \frac{1-|v_k|^2}{2}$
    for $k=0,1,2$, implying $|v_k| = \sqrt{3}^{-1}$. Since $\ket{v}$ is in the kernel of $B$, it is in the kernel of $A$ as well, which implies $\sum\limits_{k=0}^2 a_{jk} v_k = 0$ for each row. Since $|a_{jk} v_k| = 1/9$ independent of $j$ and $k$, this implies that that $(a_{jk}v_k)_{k=0,1,2}$ are three complex numbers of the same length summing to zero. They are therefore an equilateral triangle in the complex plane. Let $\sqrt{3}v = (e^{i\varphi_1},e^{i\varphi_2},e^{i\varphi_3})$, then $a_{j\cdot}$ can only be $e^{i\phi_j}\frac{1}{3}(e^{-i\varphi_1},\omega e^{-i\varphi_2},\omega^2e^{-i\varphi_3})$ or $e^{i\phi_j} \frac{1}{3}(e^{-i\varphi_1},\omega^2 e^{-i\varphi_2},\omega e^{-i\varphi_3})$ for some angle $\phi_j$. \\
    
    Now consider the matrix $C = AA^\dagger \in \C^{d_A \times d_A}$. Its non-zero eigenvalues are the same as $B$, since the Schmidt coefficients are invariant under party swap.
    Note that $C$ consists of overlaps of the two kinds of vectors described above. The overlap of two vectors of the same kind is $\frac{1}{3}$, the overlap of two vectors of different kinds is $0$. If there are $m \in \{0,...,d_1\}$ vectors of the first kind, then (after permuting rows)
    \begin{equation*}
        C = \frac{1}{3}\begin{bmatrix}1 &\cdots & 1\\\vdots & \ddots & \vdots\\\ 1& \cdots& 1 \end{bmatrix} \oplus \frac{1}{3}\begin{bmatrix}1 &\cdots & 1\\\vdots & \ddots & \vdots\\\ 1& \cdots& 1 \end{bmatrix} \in \C^{m\times m} \oplus \C^{(d_A-m)\times(d_A-m)}
    \end{equation*}
    with nonzero eigenvalues
    $\left(\frac{m}{d_1},1-\frac{m}{d_1}\right) \stackrel{!}{=} \left(\frac{1}{2},\frac{1}{2} \right)$ in contradiction to $d_A$ odd. 
\end{proof}

\begin{prop}\label{gcd prop n=2}
    Let $n=2$ and consider local Hilbert-Space dimensions $d_A\times d_B$. There always exists a group $G$ satisfying the constraints of Proposition~\ref{prop:group}. If one additionally adds the constraint, that the joint eigenbasis is not a product basis, then such a group exists, if and only if $\gcd(d_A,d_B)\neq 1$.
\end{prop}
\begin{proof} 
    Consider the generators of the group $G = \brakett{A_j\otimes B_j : j=1,...,N_G}$ (where $N_G$ is the number of generators of $G$). Since the group is abelian, they need to commute. Note that $A_j \otimes B_j$ and $A_k \otimes B_k$ commute if and only if $A_jA_k = \lambda_{j,k} A_k A_j$ and $B_j B_k = \lambda^{-1}_{j,k} B_j B_k$ for some $\lambda_{j,k} \in \C$.
    If $\lambda_{j,k} = 1$ for all choices of $j$ and $k$ then the local operators commute and therefore have a local joint eigenbasis, out of which the global one can be build. It is furthermore the only unique eigenbasis by Lemma~\ref{joint eig}.
    The choice of $\lambda_{j,k}$ is not completely free. Due to 
    \begin{equation*}
        \det(A_j)\det(A_k)=\det(A_jA_k)=\det(\lambda_{j,k} A_kA_j)=\lambda^{d_A}_{j,k}\det(A_k)\det(A_j)
    \end{equation*}
    and $|\det(A_\ell)|=1$, $\lambda_{j,k}$ has to be a $d_A$th root of unity. The same argument shows, that $\lambda^{-1}_{j,k}$ has to be a $d_B$-th root of unity (and therefore $\lambda_{j,k}$ as well). If $\gcd(d_A,d_B)=1$, then by Bézout's identity there exist $\mu,\nu \in \mathbb{Z}$ with $\mu d_A + \nu d_B = 1$ and so
    $\lambda_{j,k} = \lambda^{\mu d_A + \nu d_B}_{j,k} = (\lambda^{d_A}_{j,k})^\mu (\lambda^{d_B}_{j,k})^\nu =1$, forcing a product basis.\\
    
    For $\gcd(d_A,d_B) >1$ the construction of such a group is possible and done explicitly in Example~\ref{group const}.
\end{proof}

A similar argument for multipartite systems yields:
\begin{prop}\label{gcd prop}
    Let $n \in \mathbb{N}_{\geq 2}$ with Hilbert space dimensions $d_1,...,d_n$. If $\gcd(d_j,d_k)=1$ for every choice of $j,k \in \{1,...,n\}$ with $j\neq k$, then every group satisfying the constrains of Proposition~\ref{prop:group} has a product basis as the common eigenbasis.
\end{prop}
\begin{proof}
    Just like in Proposition~\ref{gcd prop n=2}, it is enough to show that for every pair of generators (say $M=M_1\otimes...\otimes M_n$ and $N=N_1\otimes ...\otimes N_n$) their constituent matrices commute (so $[M_j,N_j] = 0$ for all $j$). By symmetry is suffices to show that $[M_1,N_1]=0$. Let $\tilde{M} = M_2\otimes...\otimes M_n$ and $\tilde{N}=N_2\otimes...\otimes N_n$. Then $M=M_1\otimes\tilde{M}$ and $N=N_1\otimes \tilde{N}$, both being elements of $L(\C^{d_1} \otimes \C^{d_2...d_n})$. Since $\gcd(d_1,d_2...d_n)=1$ by assumption, the argument in the previous proposition shows that $M_1$ and $N_1$ (as well as $\tilde{M}$ and $\tilde{N}$) must commute.
\end{proof}

\begin{examp}[Construction of a group satisfying the constrains of Proposition~\ref{prop:group}, with a nontrivial set of eigenvectors]\label{group const}
    Let $d_A = \alpha d, d_B = \beta d$ with $\gcd(\alpha,\beta)=1$ and $d \geq 2$.

    For $d \in \mathbb{N}$ fix a computational basis $(\ket{j})_{j = 0,...,d-1}$, set $\omega_d = e^{2\pi i/d}$ and define
    \begin{equation}
        X_d = \sum_{j=0}^{d-1} \ket{j}\bra{j+1}, \quad Z_d = \sum_{j=0}^{d-1} \omega^{j}_d \ket{j}\bra{j},
    \end{equation}
    where addition inside of kets is always modulo $d$, and the subscripts are omitted if the dimension is clear.

    For $d_k = \kappa d$ ($\kappa \in \{\alpha,\beta\}, k\in \{A,B\}$) let $\tilde{Z}_k = \sum\limits_{j=0}^{d_k} \omega_{d}^j \ket{j} \bra{j} = Z_{d_k}^\kappa $ (this has order $d$)
    and consider $X_{d_A} \otimes X_{d_B}$ (this has order $\lcm(d_A,d_B) = \alpha \beta d$)
    and $\tilde{Z}_{d_A} \otimes \tilde{Z}^{-1}_{d_B}$ (of order $d$).
    Then define $G \coloneqq  \brakett{X_{d_A} \otimes X_{d_B},\tilde{Z}_A \otimes \tilde{Z}_B^{-1}}$. Note that $X_{d_A}^a \tilde{Z}_{A}^b \otimes X^a_{d_B} \tilde{Z}_{B}^{-1} \ket{1,0} = \omega_d^b\ket{a+1\; \mathrm{mod} \; d_A,a\; \mathrm{mod} \; d_B}$. Due to $X_{d_A} \tilde{Z}_{A} = \omega_d^{-1} X_{d_A} \tilde{Z}_{B}$ (same with $d_B$) the group is abelian.
    Since $(a\; \mathrm{mod} \; d_A,a\; \mathrm{mod} \; d_B)$ uniquely determines $a \; \mathrm{mod} \; \lcm(d_A,d_B)$, it follows that $G$ contains $d_Ad_B$ different elements and is isomorphic to $\mathbb{Z}_{\alpha \beta d} \times \mathbb{Z}_d$. 
    Furthermore
    \begin{equation*}
        \Tr(X_{d_A}^a Z_{d_A}^{\alpha b} \otimes X_{d_B}^a Z_{d_B}^{-\beta b}) = d_A d_B \delta_{a \in \mathbb{Z}d_A}\delta_{\alpha b \in \mathbb{Z}d_A}\delta_{a \in \mathbb{Z}d_b}\delta_{-\beta b \in \mathbb{Z}d_b} = d_A d_B \delta_{a \in \mathbb{Z} \alpha \beta b} \delta_{b \in \mathbb{Z}d},
    \end{equation*}
    so every group element (except $\mathbbm{1}$) is traceless. The eigenbasis can be explicitly given as
    \begin{equation}
        \ket{\psi_{a,b}} = \frac{1}{\sqrt{\alpha\beta d}}\sum_{\ell =0}^{\alpha \beta d-1}\omega_{\alpha \beta d}^{\ell a}\ket{ \ell \; \mathrm{mod} \; d_A, \ell +b \; \mathrm{mod} \;d_B}
    \end{equation}
    for $a \in \{0,...,\alpha \beta d-1\}$ and $b \in \{0,...,d-1\}$.
    This can be rewritten using the state \\$\ket{\phi^+} = \frac{1}{\sqrt{\alpha \beta d}} \sum\limits_{\ell = 0}^{\alpha \beta d-1} \ket{ \ell \; \mathrm{mod} \; d_A, \ell \; \mathrm{mod} \;d_B}$ and $\mu,\nu \in \mathbb{N}$ such that $\mu \alpha +\nu \beta = \gcd(\alpha,\beta)=1$ (which exist due to Bézout) as
    \begin{equation}
        \ket{\psi_{a,b}} = (Z_{d_A}^\mu\otimes Z_{d_B}^\nu)^a(\mathbbm{1} \otimes X_{d_B})^b \ket{\phi^+}.
    \end{equation}
    Then $\ket{\phi^+}$ is invariant under both group generators, and the operators in front of $\ket{\phi^+}$ in $\ket{\psi_{a,b}}$ commute with elements of $G$ up to prefactors, making $\ket{\psi_{a,b}}$ an eigenbasis. Furthermore
    \begin{equation*}
        \braket{\psi_{a,b}}{\psi_{a',b'}} = \frac{1}{\alpha \beta d} \sum_{\ell,\ell'=0}^{\alpha \beta d-1} \omega_{\alpha \beta d}^{\ell' a'-\ell a} \braket{\ell}{\ell'}_{\Hl_A}\braket{\ell+b}{\ell'+b'}_{\Hl_B}.
    \end{equation*}
    The only terms contributing in the sum fulfill $\ell-\ell' = 0 \; \mathrm{mod} \; d_A$ and $\ell -\ell ' = b'-b \; \mathrm{mod} \; d_B$.
    By the Chinese Remainder Theorem this only has a solution for $b'-b = 0 \; \mathrm{mod}  \; d$ in which case the solution for $\ell - \ell'$ is unique in $\{0,...,\alpha\beta d-1\}$ (and is given by $\ell = \ell'$). The overlap collapses to
    \begin{equation*}
        \braket{\psi_{a,b}}{\psi_{a',b'}} = \frac{\delta_{b,b'}}{\alpha \beta d} \sum_{\ell,=0}^{\alpha \beta d-1} \omega_{\alpha \beta d}^{\ell (a'-a)} = \delta_{a,a'}\delta_{b,b'}.
    \end{equation*}
    proving orthonormality.
    
    The maximally coherent states can now be written as
    \begin{equation}\label{max co G}
        \ket{\psi(\theta)} = \frac{1}{\sqrt{d_Ad_B}} \sum_{a=0}^{\alpha \beta d-1} \sum_{b = 0}^{d-1} e^{i\theta_{a,b}} \ket{\psi_{a,b}}
        = \frac{1}{\alpha \beta d^{3/2}} \sum_{a,\ell=0}^{\alpha \beta d-1} \sum_{b = 0}^{d-1} e^{i\theta_{a,b}}\omega_{\alpha \beta d}^{\ell a}\ket{ \ell \; \mathrm{mod} \; d_A, \ell +b \; \mathrm{mod} \;d_B}.
    \end{equation}
    
    Its coefficient matrix is slightly more complicated. It is given by
    \begin{equation*}
        A_{jk} = \braket{j,k}{\psi(\theta)} = \frac{1}{\alpha \beta d^{3/2}} \sum_{a,\ell=0}^{\alpha \beta d-1} \sum_{b = 0}^{d-1} e^{i\theta_{a,b}}\omega_{\alpha \beta d}^{\ell a} \delta_{j=\ell \; \mathrm{mod} \; d_A} \delta_{k = \ell +b \; \mathrm{mod} \; d_B}
    \end{equation*}
    and thus requires solving the linear congruences
    \begin{equation*}
        \ell \equiv j \mod d_A, \qquad \ell \equiv k-b \mod d_B
    \end{equation*}
    By the Chinese Remainder Theorem  this can only have a solution for $(k-b)-j \equiv 0 \mod d$, fixing $b = k-j \mod d$.
    In this case a solution exists and is unique in $\{0,...,\alpha \beta d-1\}$. Using the Bézout coefficients $\mu,\nu$, that fulfill $\mu \alpha +\nu \beta =1$, the solution can be written as
    \begin{equation*}
        \ell(j,k) = j \nu \beta+(k-[(k-j) \; \mathrm{mod} \; d]) \mu \alpha \mod \alpha \beta d
    \end{equation*}
    leading to
    \begin{equation}\label{group mat}
        A_{jk} = \frac{1}{\alpha \beta d^{3/2}} \sum_{a=0}^{\alpha \beta d-1}e^{i\theta_{a,(k-j) \; \mathrm{mod} \; d}}\omega_{\alpha \beta d}^{\ell(j,k) a}.
    \end{equation}
    
    In the case of $\beta = 1$ this expression simplifies due to $\ell(j,k) =j$.\\
\end{examp}
The $2\times2, 4\times4$ and $8\times 8$ proofs from Ref.~\cite{PRT} all first construct an orthonormal basis on the Schmidt diagonal \\$\vspan\{ \ket{kk} : k =0,...,d-1\}$, independent of the state. This approach is limited to these dimensions.
\begin{prop}\label{sym limit}
    Let $\ket{\psi} = \sum\limits_{k=0}^{d-1} c_k \ket{kk}$, for some $c_k \geq 0$. There are $d$ state-independent unitaries mapping $\ket{\psi}$ into a basis of $\vspan\{ \ket{kk} : k=0,...,d-1\}$ if and only if $d \in \{2,4,8\}$.
\end{prop}
\begin{proof}
    First note, that any element of the diagonal span is of the form $\sum\limits_{k=0}^{d-1} a_k \ket{kk}$ with Schmidt coefficients $|a_k|$. Since local unitary maps leave Schmidt coefficients invariant, they can only act as permutations and phases, so $a_k = s_{\pi(k)} e^{i\varphi_k}$. Since the initial coefficients are positive the question of existence can be rephrased as:
    Is there a set of permutations $\pi_j$ and diagonal phase matrices $S_j$, such that $\{U_jv=\pi_j S_j v : j=0,...,d\}$ are $d$ mutually orthogonal vectors for every $v \in \R^d$ with non-negative entries?\\
    
    Let $\mathcal{B}(v) = \begin{bmatrix}U_1v & \cdots & U_dv \end{bmatrix} \in \C^{d\times d}$. Then $\mathcal{B}(v)^\dagger \mathcal{B}(v) = |v|^2 \id$. At the same time, since $\mathcal{B}$ is $\R$-linear in the elements of $v$ it can be written as $\mathcal{B}(v) = \sum\limits_{j=1}^d v_j A_j$ for some complex matrices $A_j \in \C^{d\times d}$. Now $\mathcal{B}(v)^\dagger \mathcal{B}(v) = \sum\limits_{j,k=1}^d v_jv_k A_j^\dagger A_k$. Comparing coefficients in these polynomials in $(v_1,...,v_d)$ yields $A_k^\dagger A_k = \id, A_k^\dagger A_j =-A_j^\dagger A_k$ and after multiplying with $A_k$ from the left and $A_k  ^\dagger$ from the right $A_j A_k^\dagger =-A_kA_j^\dagger$.
    For $j=1,...,d-1$ let $C_j = A_j^\dagger A_d$. Then $C_j^2 =-\id$ and $C_jC_k=-C_k C_j$
    The set $\{C_j : j =1,...,d-1\}$ is thus a matrix representation of the generators of the algebra of Clifford matrices $\Cl_{d-1}(\C)$ in $\C^{d\times d}$. By the classification of Clifford representations (see \cite{Clifford} Corollary 15.30 for example) this implies $2^{\lfloor \frac{d-1}{2} \rfloor} | d$ which only happens for $d\in \{1,2,4,8\}$ (and $d=1$ is trivial).
\end{proof}

\subsection{Topological Constructions}\label{top sec}
The discussion in section~\ref{sec:group} states that universal local encodability is implied by:
\begin{equation}\label{eq:t_implies_enc}
    \big\{ \tau(B(\theta)): \theta \in \R^{d_A\times d_B} \big\} =  \mathcal{T}_{d_B}.
\end{equation}
\begin{examp}\label{d=2 examp}
    In view of (\ref{eq:t_implies_enc}) it is of interest to characterize the sets $\mathcal{T}_{d_B}$. This is simple for $d_B\in\{2,3\}$.\\
    Let $x_{\pm} = \frac{1}{3} \pm \sqrt{\frac{1}{9}- \frac{1-t_2}{6}}$, then
    \begin{equation}
    \begin{split}
        \mathcal{T}_2 &= \bigg\{ \lambda_1^2+(1-\lambda_1)^2 : \lambda_1 \in [0,1] \bigg\} = \left[\frac{1}{2},1\right], \quad
        \partial \mathcal{T}_2 = \left\{\frac{1}{2},1\right\}\\
        \mathcal{T}_3 &= \left\{(t_2,t_3) \in \left[\frac{1}{3},1 \right]\times\left[\frac{1}{9},1 \right]  : \begin{aligned}
            &\max\left\{\frac{3t_2-1}{2},2x_+^3+(1-2x_+)^3\right\}\leq t_3,\\
            &\qquad \qquad t_3 \leq2x_-^3+(1-2x_-)^3
        \end{aligned} \right\}
    \end{split}
    \end{equation}
    and $\partial \mathcal{T}_3$ is given by the union of the graphs of $t_3(t_2) = \frac{3t_2-1}{2}$ (one eigenvalue is zero) for $t_2 \in \left[ \frac{1}{2},1 \right]$, $t_3(t_2) = 2x_+^3+(1-2x_+)^3$ for $t_2 \in \left[ \frac{1}{3},\frac{1}{2} \right]$ (the bigger two eigenvalues are identical) and $t_3(t_2) = 2x_-^3+(1-2x_-)^3$ for $t_2 \in \left[ \frac{1}{3},1 \right]$ (the smallest two eigenvalues are identical).
\end{examp}
\begin{figure}[ht!]
	\begin{center}

    \begin{tikzpicture}[scale=10]
        \draw[help lines] (0,0) grid [step=1/10] (1.1,1.1);
        \draw[-latex] (0,0) -- (1.1,0) node [right] {$\tr(B^2)$};
        \draw[-latex] (0,0)     -- (0,1.1) node [above] {$\tr(B^3)$};
        \foreach \i in {0,0.5,1}
          \draw (\i,0.02) -- (\i,-0.02) node [below] {$\i$};
        \foreach\i in {0,0.5,1}
          \draw (0.02,\i) -- (-0.02,\i) node [left]  {$\i$};
        \draw[red, thick, name path = A] plot[variable=\t,domain=0:1/2,smooth,thick] ({2*\t^2+(1-2*\t)^2},{2*\t^3+(1-2*\t)^3});
        \draw[red, thick, name path = B] plot[variable=\t,domain=0:1,smooth,thick] ({\t^2+(1-\t)^2},{\t^3+(1-\t)^3});
        \tikzfillbetween[of=A and B, on layer = main]{red};
        \node [left] at (1/3,1/9) {$(\frac{1}{3},\frac{1}{9})$};
        \node [left] at (1,1) {$(1,1)$};
        \node [right] at (1/2,1/4) {$(\frac{1}{2},\frac{1}{4})$};
        \node at (0.75,0.5) {$\Gamma_1$};
        \node at (0.45,0.13) {$\Gamma_2$};
        \node at (0.46,0.34) {$\Gamma_3$};
    \end{tikzpicture}
	\end{center}
	\caption{The set $\mathcal{T}_3$ displayed in red. Its boundary is naturally divided into three parts, denoted by $\Gamma_j$. Note that $\mathcal{T}_3$ looks vaguely triangular, though only the path $\Gamma_1$ between $(1,1)$ and $(\frac{1}{2},\frac{1}{4})$ is truly a straight line. It can easily be seen that this set is compact, contractible and homeomorphic to a ball (which is true for all $\mathcal{T}_d$, see Proposition~\ref{T_d char}). The corner at $\left(\frac{1}{2},\frac{1}{4}\right)$ corresponds to the state $\frac{1}{\sqrt{2}}(\ket{00}+\ket{11})$ from Remark~\ref{LME rem}.}
\end{figure}
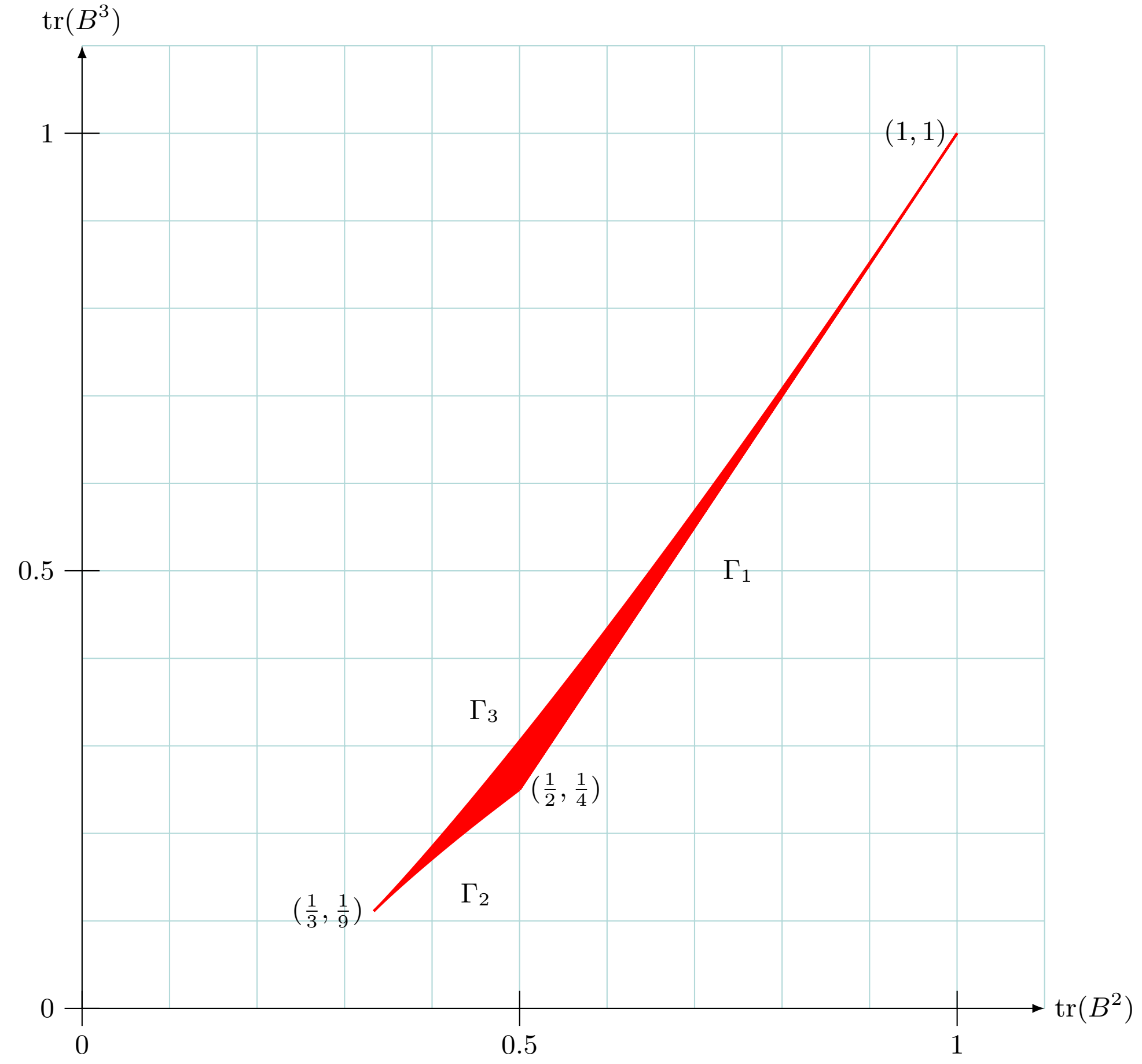

\begin{prop}\label{T_d char}
    Fix $d \in \N$. $\mathcal{T}_d$ is compact, contractible and its boundary is given by the trace vectors, resulting from (at least) two equal eigenvalues or one zero eigenvalue. Furthermore $\mathcal{T}_d$ is homeomorphic to $\overline{B_1(0)}$ where the homeomorphism maps $\partial \mathcal{T}_d$ to $S^{d-2}$.
\end{prop}
\begin{proof}
    Consider the set compact, convex and contractible set
    \begin{equation*}
        K = \{ x \in \R^{d-1} : 0 \leq x_1 \leq x_2 \leq ... x_{d-1}\leq 1-x_1-...-x_{d-1} \}.
    \end{equation*}
    Its boundary is given by $\{ x\in K \;|\; \exists j \in \{0,...,d\} : x_j = x_{j+1}\}$ with $x_0=0$, $x_d =1-x_1-...-x_{d-1}$ and $x_{d+1} = 1$. Let $f(x) = \left( \,\sum\limits_{j=1}^d x_j^2,...,\sum\limits_{j=1}^d x_j^d\right)$. By Newtons identities this function is injective, and $\mathcal{T}_d = f(K)$. Since $K$ is compact and $f$ is continuous and injective, it is a homeomorphism onto its image. This extends to its boundary. The claimed boundary for $\mathcal{T}_d=f(K)$ is $f(\partial K)$. It remains to show $f(\partial K) = \partial f(K)$:\\

    Let $x \in K^\circ$, then there is an open region $U \subset K$, such that $x \in U$. By invariance of domain $f(U) \subset f(K)$ is open in $\R^{d-1}$, and thus $y = f(x) \in \mathcal{T}^\circ_d$. This shows $f(K^\circ) \subseteq\mathcal{T}^\circ_d $.
    
    Now fix $y \in \mathcal{T}^\circ_d$. There is an open set $y \in U' \subset \mathcal{T}_d$. Let $g : \mathcal{T}_d \to K \subset \R^n$ be the continuous (bijective) inverse of $f : K \to f(K)$. Again by invariance of domain $g(U') \subset K$ is open, since $g$ is injective, when viewed as a map to $\R^n$, and $x = g(y) \in g(U')$, thus $x \in K^\circ$, implying $g(\mathcal{T}_d^\circ)\subseteq K^\circ$ and thus $\mathcal{T}_d^\circ \subseteq f(K^\circ)$. Combining the two inclusions leads to equality. Since $K$ and $f(K)=\mathcal{T}_d$ are compact, their boundary is given by $\partial K =K \backslash K^\circ$ and 
    \begin{equation*}
        \partial f(K)=f(K) \backslash f(K)^\circ = f(K) \backslash f(K^\circ) = f(K\backslash K^\circ) = f(\partial K),
    \end{equation*}
    where the second to last step used injectivity.

    The last claim follows directly, from the fact, that any convex set with nonempty interior (in particular $K$) is homeomorphic to the unit ball, with its boundary being mapped to the sphere.
\end{proof}
The reason to consider these properties of $\mathcal{T}$ is to generalize the following result:
\begin{prop}\label{dx2}
    Every state is locally encodable for two parties and local dimension $d_A \geq 2$ and $d_B=2$.
\end{prop}
\begin{proof}
    First assume, there are sets of angles $\theta^{(1)},\theta^{(2)} \in \R^{d_A \times 2}$, such that $\ket{\psi(\theta^{(1)})}$ is a product state, and $\ket{\psi(\theta^{(2)})}$ is maximally mixed. Now consider the set of states $\ket{\psi(t\theta^{(1)}+(1-t)\theta^{(2)})}$ for $t\in \left[0, 1\right]$ and the set
    \begin{equation*}
        \left\{ \tau(B(t\theta^{(1)}+(1-t)\theta^{(2)})) : t\in \left[0, 1\right] \right\} \subseteq \mathcal{T}_2
    \end{equation*}
    Since the map $t \mapsto \tau(B(t\theta^{(1)}+(1-t)\theta^{(2)}))$ is continuous, this is a connected subset of $\mathcal{T}_2$. Since $\mathcal{T}_2 = \left[ \frac{1}{2},1 \right]$ and the set includes both endpoints, it is already equal to $\mathcal{T}_2$.
    Such angles do exist: For example use the cyclic construction from example~\ref{prod ex} and the angles $\theta^{(1)} =0, \theta^{(2)}_{jk} = \omega_{d}^{jk}$. They do also exist for the construction from example~\ref{group const} (for even $d_A$) by again setting $\theta^{(1)} =0$ and now $\theta^{(2)}_{jk} = \frac{\pi}{4} \delta_{j \neq k}$ (for $d_A=2$) and $\theta_{j,k}^{(2)} =  \frac{4\pi j\delta_{k,1}}{d_A}$ (for $d_A \geq 4$).
\end{proof}

\begin{prop}\label{angle function}
    Let $d_A, d_B \in \N, \xi : S^{d_B-2} \to \R^{d_A\times d_B}$ be continuous, such that
    \begin{equation*}
        \tilde{\Phi} : S^{d_B-2} \to \partial \mathcal{T}, \quad t \mapsto \tau(B(\xi(t)))
    \end{equation*}
    is well defined and not homotopic to the constant function. Then $\{\tau(B(\theta)) : \theta \in \R^{d_A \times d_B}\} = \mathcal{T}_{d_B}$.
\end{prop}
\begin{proof}
    Let $g : \mathcal{T}_{d_B} \to \overline{B_1(0)} \subseteq \R^{d_B-1}$ be the homeomorphism obtained by Proposition~\ref{T_d char} with $g(\partial \mathcal{T}) = S^{d_B-2}$.
    Let $f = g \circ \tau \circ B : \R^{d_A\times d_B} \to \overline{B_1(0)}$. The condition can now be reformulated as $f \circ \xi$ is not homotopic to the constant function and maps into $S^{d_B-2}$.
    This already implies that $f \circ \xi$ is surjective, since otherwise the maps takes values in a punctured sphere, which is contractible (and thus $f\circ \xi$ null-homotopic).

    Assume that $f$ is not surjective (which corresponds to a point in $\mathcal{T}_{d_B}$ which is not reached by any $\tau(B(\theta))$), then the image of $f$ is missing at least one point in the inside of $\overline{B_1(0)}$. Therefore there exists a retraction $r$, such that $r \circ f : \R^{d_A\times d_B} \to S^{d_B-2}$. Consider $H(x,t) = (1-t) \xi(x)$, which continuously deforms $\xi$ to $0$. Thus $r\circ f \circ H$ deforms $r\circ (f\circ \xi)$ to the constant map. But $f \circ \xi$ lies on $S^{d_B-2}$, thus $r\circ f \circ \xi = f\circ \xi$, which is not null-homotopic by assumption. 
\end{proof}
Such a map will be constructed for $d_A=d_B=3$ in the following chapter.
\section{The explicit construction for $3\times 3$}\label{3x3}
\subsection{The $A$ and $B$ Matrices and the Fourier-Transform for $d_A=d_B=3$}
For $d_A=d_B=3$ the matrix from Equation (\ref{group mat}) takes the form 
    \begin{equation}\label{A thet}
        A(\theta)= \frac{1}{3\sqrt{3}} \begin{bmatrix} e^{i \theta_{0,0}} + e^{i \theta_{1,0}} + e^{i \theta_{2,0}} & e^{i \theta_{0,1}} + e^{i \theta_{1,1}} + e^{i \theta_{2,1}} & e^{i \theta_{0,2}} + e^{i \theta_{1,2}} + e^{i \theta_{2,2}} \\
        e^{i \theta_{0,2}} + \omega e^{i \theta_{1,2}} + \omega^2 e^{i \theta_{2,2}} & e^{i \theta_{0,0}} + \omega e^{i \theta_{1,0}} + \omega^2 e^{i \theta_{2,0}} & e^{i \theta_{0.1}} + \omega e^{i \theta_{1,1}} + \omega^2 e^{i \theta_{2,1}} \\
        e^{i \theta_{0,1}} + \omega^2 e^{i \theta_{1,1}} + \omega e^{i \theta_{2,1}} & e^{i \theta_{0,2}} + \omega^2 e^{i \theta_{1,2}} + \omega e^{i \theta_{2,2}} & e^{i \theta_{0,0}} + \omega^2 e^{i \theta_{1,0}} + \omega e^{i \theta_{2,0}}
        \end{bmatrix}
    \end{equation}
    Choosing $\begin{bmatrix}
        \theta_{0,2} \\ \theta_{1,2} \\ \theta_{2,2}
    \end{bmatrix} = \begin{bmatrix}
        \theta_{0,1} \\ \theta_{1,1} - \frac{2\pi}{3} \\ \theta_{2,1} + \frac{2\pi}{3}
    \end{bmatrix}$ makes $A(\theta)$ symmetric.
    Let $\mathcal{F} = \frac{1}{\sqrt{d}} \left(\omega^{jk} \right)_{j,k = 0,...,d-1}$
    be the DFT matrix and set $x = \mathcal{F} \begin{bmatrix}e^{i\theta_{0,0}} \\ e^{i \theta_{1,0}} \\ e^{i \theta_{2,0}} \end{bmatrix},w = \mathcal{F} \begin{bmatrix}e^{i\theta_{0,1}} \\ e^{i \theta_{1,1}} \\ e^{i \theta_{2,1}} \end{bmatrix}$, such that $A = \frac{1}{3} \begin{bmatrix} x_0 & w_0 & w_2 \\ w_0 & x_1 & w_1 \\ w_2 & w_1 & x_2 \end{bmatrix}$ and    
    \begin{equation}\label{B eq}
        B = A^\dagger A = \frac{1}{9} \begin{bmatrix} |x_0|^2+|w_0|^2+|w_2|^2 & x_0^*w_0+w_0^*x_1+w_2^*w_1 & x_0^*w_2 + w_0^*w_1+w_2^*x_2 \\
        w_0^*x_0+x_1^*w_0+w_1^*w_2 & |w_0|^2+ |x_1|^2+|w_1|^2 & w_0^*w_2+x_1^*w_1+w_1^*x_2 \\ w_2^*x_0+w_1^*w_0+x_2^*w_2 & w_2^*w_0+w_1^*x_1+x_2^*w_1 & |w_2|^2 + |w_1|^2 +|x_2|^2
        \end{bmatrix}.
    \end{equation}
    In the following matrices of the above form are constructed, without first stating an explicit closed form for the angles. It is therefore important to know, when a given matrix $\tilde{A} \in \C^{3\times 3}$ can be written in this form for some choice of $\theta \in \R^{3\times 3}$. Under the symmetric choice of angles, there is an easy condition to check.
    
    \begin{lemma}\label{Fourier cond}
        Let $x \in \C^3$ be given. There are angles $\theta_0, \theta_1, \theta_2$ such that $x =  \mathcal{F} \begin{bmatrix}e^{i\theta_{0}} \\ e^{i \theta_{1}} \\ e^{i \theta_{2}} \end{bmatrix}$ if and only if $||x||_2^2 = 3$ and $x_0 x_1^* + x_1 x_2^*+x_2x_0^* = 0$
    \end{lemma}
    \begin{proof}
        \underline{Necessity:} Since $\mathcal{F}$ is a unitary map, $||x||_2^2 = || \mathcal{F}^{-1}x||_2^2 = 3$. Furthermore 
        \begin{equation*}
        \begin{split}
            \sum_{j=0}^{2} x_j x_{j+1}^* &= \frac{1}{3} \sum_{j,k,\ell=0}^{2} \omega^{jk} e^{i \theta_k} \omega^{-(j+1)\ell} e^{-i \theta_{\ell}} = \sum_{k,\ell=0}^{2}\omega^{-\ell} e^{i \theta_k - \theta_\ell} \underbrace{\sum_{j=0}^{2} (\omega^{k-\ell})^j}_{=3 \delta_{k,\ell}} = 3 \sum_{k,\ell=0}^{2} \delta_{k,\ell} \omega^{-\ell} e^{i \theta_k - \theta_\ell} = 3\sum_{\ell = 0}^2 \omega^{-\ell} = 0.
        \end{split}
        \end{equation*}
        \underline{Sufficiency:} Let $\alpha = \mathcal{F}^{-1}x$. It needs to be shown, that $|\alpha_j|^2 = 1$ for all $j\in \{0,1,2\}$. This follows from
        
        \begin{equation*}
            |\alpha_j|^2 = \frac{1}{3} \left|\sum_{\ell = 0}^2 \omega^{-j\ell}x_\ell \right|^2 = \frac{1}{3} \sum_{\ell,m = 0}^2 \omega^{j(m-\ell)}x_m x_\ell^* = \frac{1}{3}\sum_{k=0}^2 \omega^{-jk} \underbrace{\sum_{m = 0}^2 x_m x_{k+m}^*}_{=3\delta_{k = 0}} = \omega^{0} = 1,
        \end{equation*}
        since the three possible terms in the $m$ sum are the norm of $x$ ($k=0$), the constraint on $x$ ($k=1$) and its complex conjugate ($k=2$).
    \end{proof}
    In the following $\partial\mathcal{T}_3$ is split into the three parts $\Gamma_1,\Gamma_2$ and $\Gamma_3$, joint at the extremal points \\$(t_2,t_3) \in \{(1,1),(1/2,1/4),(1/3,1/9)\}$. A continuous function $\xi$ as is required by Proposition~\ref{angle function} is constructed on each $\Gamma_j$ separately, and then proven to periodic, when joined up. Note that this $\xi$ then cannot be homotopic to a point path, if it is bijective on one of the three boundary parts.
\subsection{Boundary $\Gamma_1$: One eigenvalue vanishes}\label{B1}
    An eigenvalue of $B$ vanishes automatically, if two of the rows of $B$ repeat. This can be forced via 
    \begin{equation*}
        w_0 = x_0 = x_1, \quad w_2 = w_1.
    \end{equation*}
    If $x_0$ is furthermore assumed to be real, then the conditions of Lemma~\ref{Fourier cond} become
    \begin{equation*}
        2x_0^2+|x_2|^2 = 3, \quad x_0^2+2x_0\Real(x_2) = 0, \quad x_0^2 + 2 |w_1|^2 = 3, \quad 2x_0 \Real(w_1)+|w_1|^2 = 0,
    \end{equation*}
    which can be solved for $x_0 \in [1/\sqrt{3},2/\sqrt{3}]$ yielding
    \begin{equation}\label{concr x}
        x = \begin{bmatrix} x_0 \\x_0 \\ - \frac{x_0}{2} \pm \frac{3i}{2}\sqrt{\frac{4}{3}-x_0^2} \end{bmatrix}, \quad w= \begin{bmatrix} x_0 \\ \frac{x_0^2-3}{4x_0} \pm \frac{3i}{4x_0} \sqrt{(3-x^2_0)\left(x^2_0-\frac{1}{3}\right)}\\ \frac{x_0^2-3}{4x_0} \pm \frac{3i}{4x_0} \sqrt{(3-x^2_0)\left(x^2_0-\frac{1}{3}\right)}\end{bmatrix},
    \end{equation}
    where the sign of $x_2$ and the signs of $w_2=w_3$ are independent. A sign change in both $x_2$ and $w_2=w_3$ corresponds to a complex conjugation of $A$ and therefore $B$, which only results in the conjugation of the (real valued) eigenvalues. Thus the sign in $x_2$ can be fixed to be $+$.
    
    The two signs for the imaginary parts of $w_1=w_2$ correspond to two continuous families of matrices, whose trace-vectors lie on $\partial \mathcal{T}_3$. Due to $w_1\left(x_0=\frac{1}{\sqrt{3}}\right) \in \R$ the two families can be continuously joined.
    Note that Equation (\ref{B eq}) now becomes
    \begin{equation}
        B=\frac{1}{9}\begin{bmatrix}
            \alpha & \alpha & \beta \\ \alpha & \alpha & \beta \\ \beta^* & \beta^* & \gamma
        \end{bmatrix} \quad \text{with} \quad \begin{bmatrix} \alpha \\\beta \\ \gamma \end{bmatrix} = \begin{bmatrix} 2|x_0|^2+|w_1|^2\\2x_0w_1+w_1^*x_2 \\ 2|w_1|^2+|x_2|^2\end{bmatrix}.
    \end{equation}
    The characteristic polynomial of $B$ is given by
    \begin{equation*}
        \chi_{B}(\lambda) = - \lambda^3 + \lambda^2\frac{2\alpha+\gamma}{9}-2\lambda\frac{\alpha\gamma - |\beta|^2}{81} = - \lambda^3 + \lambda^2-\lambda\lambda_0(1-\lambda_0)
    \end{equation*}
    where the eigenvalues are $\lambda_0,1-\lambda_0$ and $0$. The coefficient in front of $\lambda$ is given by $(\lambda_0(1-\lambda_0)) \in \left[0, \frac{1}{4}\right]$ and determines $\lambda_0$ uniquely (under the condition $\lambda_0 \leq 1 - \lambda_0$).
    Thus it remains to show, that $\alpha\gamma - |\beta|^2$ as a function of $x_0$ can obtain any possible value in $[0, \frac{81}{8}]$. Note that only $\beta$ depends on the choice of sign. Furthermore a short calculation shows
    \begin{equation*}
        (\alpha\gamma-|\beta_+|^2)\left(x_0=\frac{1}{\sqrt{2}}\right) =  \frac{81}{8}, \qquad (\alpha\gamma-|\beta_-|^2)(x_0=1)=0.
    \end{equation*}
    which is sufficient by the mean value theorem, since the two branches can be continuously joined at $x_0=\frac{1}{\sqrt{3}}$. More precisely, the $-$ branch starts at $x_0=1$ which corresponds to $\lambda_0=0$ and decreases $x_0$ to $\frac{1}{\sqrt{3}}$ where it is joined with the $+$ branch which then increases $x_0$ to $\frac{1}{\sqrt{2}}$ which corresponds to $\lambda_0 = \frac{1}{2}$.
    
    The interpolation in Lemma~\ref{angle function} interpolates between angles, not points on the sphere. Since $e^{ia}$ and $e^{ib}$ get interpolated differently than $e^{ia}$ and $e^{i(b+2\pi)}$ (even though they are the same numbers), the angles need to be fixed. The idea is to construct three continuous functions of angles $\xi_j$, such that $\tau(B(\xi_j))$ parametrizes one of the three boundaries of $\mathcal{T}_3$ and then join these three. For this they need to all be continuous and equal at their overlapping points. For $z \notin (-\infty,0]$ define $\arg(z) $ as the unique $\varphi \in (-\pi,\pi)$ such that $z=|z|e^{i \arg(z)}$. Now
\begin{equation*}
\begin{split}
    \begin{bmatrix}
        e^{i\theta_{00}} \\ e^{i \theta_{10}} \\ e^{i \theta_{20}}
    \end{bmatrix} &= \mathcal{F}^{-1} x = \mathcal{F}^{-1} \begin{bmatrix} x_0 \\ x_0 \\ -\frac{x_0}{2} + \frac{3i}{2}\sqrt{\frac{4}{3}-x_0^2}\end{bmatrix} = \begin{bmatrix} \frac{\sqrt{3}}{2}\left(x_0+i\sqrt{\frac{4}{3}-x_0^2} \right) \\ \frac{\omega\sqrt{3}}{2}\left(-x_0+i\sqrt{\frac{4}{3}-x_0^2} \right)\\\frac{\omega^2\sqrt{3}}{2}\left(-x_0+i\sqrt{\frac{4}{3}-x_0^2} \right)\end{bmatrix} \\
    \begin{bmatrix}
        e^{i\theta_{01}^{\pm}} \\ e^{i \theta_{11}^{\pm}} \\ e^{i \theta_{21}^{\pm}}
    \end{bmatrix} &= \mathcal{F}^{-1} w^{\pm} = \mathcal{F}^{-1} \begin{bmatrix} x_0 \\ \frac{x_0^2-3}{4x_0}\pm \frac{3i}{4x_0} \sqrt{(3-x_0^2)\left(x_0^2-\frac{1}{3}\right)} \\ \frac{x_0^2-3}{4x_0}\pm \frac{3i}{4x_0} \sqrt{(3-x_0^2)\left(x_0^2-\frac{1}{3}\right)}\end{bmatrix} = \frac{1}{\sqrt{3}} \begin{bmatrix} x_0 + 2w_1^{\pm} \\x_0-w_1^{\pm} \\ x_0-w_1^{\pm} \end{bmatrix}
\end{split}
\end{equation*}

This can be continuously achieved with
\begin{equation*}
\begin{split}
    \theta_{00} &= \arg\left(\frac{\sqrt{3}}{2}\left(x_0+i\sqrt{\frac{4}{3}-x_0^2} \right)\right),\quad \theta_{10} = \frac{5\pi}{3} - \theta_{00}, \quad \theta_{20} = \theta_{10} + \frac{2\pi}{3} \\
    \theta_{01}^{+} &= \arg\left(\frac{x_0+2w_1^+}{\sqrt{3}}\right), \quad\theta_{01}^{-} = \arg\left(\frac{x_0+2w_1^-}{\sqrt{3}}\right)+ 2\pi, \quad \theta_{11}^\pm = \theta_{21}^\pm = \arg\left(\frac{x_0-w_1^\pm}{\sqrt{3}}\right)
\end{split}
\end{equation*}
Note, that $\theta_{00}(x_0)$ is continuous due to $x_0 > 0$.
$\theta_{01}^{\pm}$ and $\theta_{11}^{\pm}$ are continuous, since their imaginary part is proportional to $\sqrt{(3-x_0^2)\left(x_0^2-\frac{1}{3}\right)}$, which is nonzero for $x_0 \in \left(\frac{1}{\sqrt{3}},1\right)$. It is zero for $x_0 = \frac{1}{\sqrt{3}}$, where $\theta_{01}^{\pm} = -1$ and $\theta_{11}^{\pm}= 1$. Thus $\theta_{11}^{\pm}$ is continuously well defined, while $\theta_{01}^{\pm}$ must be continuously extended. Since $\Imag(\theta_{01}^{+}) \geq 0$ its continuous extension at $x_0 = \frac{1}{\sqrt{3}}$ is given by $\pi$, while $\arg\left(\frac{x_0+2w_1^-}{\sqrt{3}}\right)$ has to be extended by $-\pi$. For $\theta_{01}^+$ and $\theta_{01}^-$ to agree at $x_0 = \frac{1}{\sqrt{3}}$, $\theta_{01}^-$ is shifted by $+2\pi$.
\subsection{Boundary $\Gamma_2$: From $\lambda =(1/2,1/2,0)$ to $\lambda =(1/3,1/3,1/3)$ - The two bigger eigenvalues are identical}\label{B2}
Let $\varphi_0 = \arccos(\sqrt{6}/4)$, $\theta_t = (1-t) \arccos(1/4)$ and set
\begin{equation}
    x = e^{i\varphi_0} \mathcal{F} \begin{bmatrix}1 \\ \omega e^{i\theta_t} \\ \omega^2 e^{i \theta_t} \end{bmatrix} = \frac{e^{i\varphi_0}}{\sqrt{3}} \begin{bmatrix} 1-e^{i \theta_t} \\ 1-e^{i \theta_t} \\ 1+2e^{i\theta_t} \end{bmatrix}, \quad w = e^{i\varphi_0} \mathcal{F} \begin{bmatrix} e^{i\theta_t} \\ e^{-i\theta_t} \\ e^{-i\theta_t} \end{bmatrix} = \frac{e^{i\varphi_0}}{\sqrt{3}} \begin{bmatrix} e^{i \theta_t}+2e^{-i \theta_t} \\ e^{i \theta_t}- e^{-i \theta_t} \\ e^{i \theta_t}- e^{-i \theta_t} \end{bmatrix}.
\end{equation}
In particular $x_0 = x_1$ and $w_1 = w_2$. Note that
\begin{equation*}
\begin{split}
    B_{31} = B_{32}  & = \frac{w_1^*(x_0+w_0)+x_2^*w_1}{9}\\
    &= \frac{1}{27}\left[(e^{-i\theta_t}-e^{i\theta_t})(1-e^{i\theta_t}+e^{i\theta_t}+2e^{-i\theta_t})+(1+2e^{-i\theta_t})(e^{i\theta_t}-e^{-i\theta_t})\right] = 0,
\end{split}
\end{equation*}
implying $B$ is of the form
$\begin{bmatrix}
    B_{11} & B_{12} & 0 \\ B_{12} & B_{22} & 0 \\ 0 & 0& B_{33},
\end{bmatrix}$
and its eigenvectors are $\begin{bmatrix} 1 & \pm 1 & 0 \end{bmatrix}^{\T}$ and $\begin{bmatrix} 0 & 0 &1 \end{bmatrix}^{\T}$ with eigenvalues $B_{11} \pm B_{12}$ and $B_{33}$.
Due to
\begin{equation*}
\begin{split}
    B_{33}  &= \frac{2|w_2|^2+|x_2|^2}{9} = \frac{1}{27}(2|e^{i\theta_t}-e^{-i\theta_t}|^2+ |1+2e^{i\theta_t}|^2)\\
    &= \frac{9+4\cos(\theta_t)-4\cos(2\theta_t)}{27} = \frac{|x_0|^2+|w_0|^2+2|w_2|^2+2\Real(w_0^*x_0)}{9} = B_{11} + B_{21}\\
\end{split}
\end{equation*}
this one-parameter family of matrices always has a repeated eigenvalue. This eigenvalue lies in $\left[\frac{1}{3},\frac{1}{2}\right]$, since $f(\theta) = 9 + 4\cos(\theta)-4\cos(2\theta)$ is strictly monotonically increasing for $\theta \in [0, \arccos(1/4)]$ (its derivative is $8 \sin(x/2) \cos(x/2) (4 \cos(x)-1) \geq 0$) and $B_{33}(t=0) = \frac{1}{2}$ and $B_{33}(t=1) =\frac{1}{3}$.
The corresponding angles now induce a parametrization of $\Gamma_2$. Furthermore, since $f(\theta)$ is strictly monotonic, the angle function is a bijective map from $[0,1]$ onto $\Gamma_2$.

\subsection{Boundary $\Gamma_3$: From $\lambda =(1/3,1/3,1/3)$ to $\lambda =(1,0,0)$ - The smaller two eigenvalues are identical}\label{B3}
With $\varphi_0$ as before, let $\varphi_t = (1-t)\varphi_0+ t\pi/6, \;\phi_t = t2\pi/3$ and set

\begin{equation}
    x = e^{i\varphi_t} \mathcal{F} \begin{bmatrix}1 \\ \omega e^{i\phi_t} \\ \omega^2 e^{i \phi_t} \end{bmatrix} = \frac{e^{i\varphi_t}}{\sqrt{3}} \begin{bmatrix} 1-e^{i \phi_t} \\ 1-e^{i \phi_t} \\ 1+2e^{i\phi_t} \end{bmatrix}, \quad w = e^{i\varphi_t} \mathcal{F} \begin{bmatrix} e^{2i\phi_t} \\ 1 \\ 1 \end{bmatrix} = \frac{e^{i\varphi_t}}{\sqrt{3}} \begin{bmatrix} 2+e^{2i\phi_t} \\ e^{2i\phi_t} -1 \\ e^{2i\phi_t} -1 \end{bmatrix}
\end{equation}
Notice that $x$ and $w$ at $t=0$ are constructed to equal the $x$ and $w$ at $t=1$ from Boundary 2. So this paths starts at $(1/3,1/3,1/3)$ and with compatible angles.
Explicitly calculating entries of $B$ yields:

\begin{equation}\label{expl B}
\begin{split}
    B_{31} &= \frac{w_1^*(x_0+w_0)+x_2^*w_1}{9} =\frac{e^{-2i\phi_t}-e^{-i\phi_t}-1+e^{i\phi_t}}{9} \Rightarrow |B_{31}|^2 =\frac{16}{81} \sin^2(\phi_t/2) \sin^2(\phi_t)\\
    B_{33} &= \frac{2|w_2|^2+|x_2|^2}{9} = \frac{9+4\cos(\phi_t)-4\cos(2\phi_t)}{27} \\
    B_{12} &= \frac{2\Real(x_0^*w_0)+|w_1|^2}{9} = \frac{2}{9} (1-\cos(\phi_t)). \\
\end{split}
\end{equation}
Due to $x_0 = x_1$ and $w_1 = w_2$, $\begin{bmatrix} 1 & -1 & 0 \end{bmatrix}^{\T}$ is an eigenvector of $B$ with eigenvalue
\begin{equation*}
    B_{11}-B_{12} = \frac{|x_0|^2+|w_0|^2-2 \Real(x_0^*w_0)}{9} =\frac{(2 \cos(\phi_t)+1)^2}{27},
\end{equation*}
which is monotonically decreasing for $\phi_t \in [0, \frac{2\pi}{3}]$ going between $1/3$ and $0$.

And indeed this eigenvalue is repeated as well. To prove this, it should be noted at first, that by (\ref{expl B}) $B_{31}$ is never zero for $\phi_t \in (0,\frac{2\pi}{3}]$, making division by $B_{31} = B_{13}^*$ well defined for $t \neq 0$. Furthermore consider now an eigenvector of the Form $v_\chi = \begin{bmatrix} 1 & 1 &  \chi \end{bmatrix}^{\T}$ for some $\chi \in \C$. $v_\chi$ being an eigenvector to the eigenvalue $B_{11}-B_{12}$ is equivalent to
\begin{equation*}
    \begin{bmatrix}
    B_{11} + B_{12} + \chi B_{13}\\
    2 B_{13}^* + \chi B_{33}
    \end{bmatrix} \stackrel{!}{=}  \begin{bmatrix} B_{11}-B_{12} \\ \chi (B_{11} - B_{12}) \end{bmatrix}.
\end{equation*}
The first equation forces $\chi = - \frac{2B_{12}}{B_{13}}$ which leads to the condition $|B_{13}|^2  \stackrel{!}{=}  B_{12} (B_{33}-(B_{11}-B_{12}))$,
which is true because trigonometric identities yield
\begin{equation*}
    \frac{16}{81} \sin^2(\phi_t/2)\sin^2(\phi_t) = \frac{2(1-\cos(\phi_t))}{9\cdot 27}(9+4\cos(\phi_t)-4\cos(2\phi_t)-(3+4\cos(\phi_t)+2\cos(2\phi_t))).
\end{equation*}
Thus $B_{11}-B_{12}$ is indeed an eigenvalue for the linearly independent vectors $\begin{bmatrix}1 & -1 & 0 \end{bmatrix}^{\T}$ and $\begin{bmatrix}1 &1 & - \frac{2B_{12}}{B_{13}} \end{bmatrix}^{\T}$. Due to $(B_{11}-B_{12})(t=1) = 0$ the angle function spans the entirety of $\Gamma_3$.

\subsection{Comparing Angles at Endpoints}\label{Compare}
The beginning of the angle function on $\Gamma_1$ and the end of the angle function on $\Gamma_3$ are both at the eigenvalues $(1,0,0)$ achieved via the angles

\begin{equation*}
    \theta_{00}  = \frac{\pi}{6}, \quad \theta_{10} = \frac{3\pi}{2}, \quad\theta_{20} = 2\pi +\frac{\pi}{6}, \quad\theta_{01} = \frac{3\pi}{2}, \quad\theta_{11} = \theta_{21}  = \frac{\pi}{6}
\end{equation*}

The end of the construction on $\Gamma_1$ and the beginning of the construction on $\Gamma_2$ both correspond to the eigenvalues $\left(\frac{1}{2},\frac{1}{2},0\right)$ achieved via the angles
\begin{equation*}
\begin{split}
    \theta_{00} &= \arccos\left(\frac{\sqrt{6}}{4}\right), \quad\theta_{10} = \frac{5\pi}{3} - \arccos\left(\frac{\sqrt{6}}{4}\right) = \arccos\left(\frac{\sqrt{6}}{4}\right)+\arccos\left(\frac{1}{4}\right)+\frac{2\pi}{3}, \quad\theta_{20} = \theta_{10} + \frac{2\pi}{3}, \\
    \theta_{01} &= \arg\left(\frac{-\sqrt{6}+i\sqrt{10}}{4}\right) = \arccos\left(\frac{-\sqrt{6}}{4}\right), \quad\theta_{11} = \theta_{21} = \arg\left(\frac{3\sqrt{6}-i\sqrt{10}}{8}\right) = - \arccos\left(\frac{3\sqrt{6}}{8}\right)
\end{split}
\end{equation*}

The $\Gamma_2$ construction ends at eigenvalues $\left(\frac{1}{3},\frac{1}{3},\frac{1}{3}\right)$, meeting the $\Gamma_3$ construction with the angles
\begin{equation*}
    \theta_{00} = \arccos\left(\frac{\sqrt{6}}{4}\right), \quad \theta_{10} = \theta_{00}+ \frac{2\pi}{3}, \quad \theta_{20} = \theta_{00}+ \frac{4\pi}{3}, \quad
    \theta_{01} = \theta_{11} = \theta_{21} = \theta_{00}
\end{equation*}

\subsection{Conclusion}
\begin{theorem}\label{qutrit possible}
    Let $\mathcal{H}_A,\mathcal{H}_B$ be two complex Hilbert spaces of dimension $3$. Then for any state $\ket{\psi} \in \mathcal{H}_A \otimes \mathcal{H}_B$, the set $\orb_{\LU}(\ket{\psi})$ contains an orthonormal basis. In short: Universal local encodability is possible for two qutrits.
\end{theorem}
\begin{proof}
    Sections~\ref{B1} to~\ref{B3} have constructed continuous functions $\xi_j :I \to \R^{3\times 3}$, such that $\Image(\tau \circ B \circ \xi_j) = \Gamma_j$ (where $I$ is some interval). Section~\ref{Compare} shows, that $\xi_1,\xi_2,\xi_3$ can be combined into one periodic function $\xi$, since they all agree at their respective points of overlap. Since $\xi_2$ (and $\xi_3$) is bijective, $\xi$ cannot be homotopic to a point path on $\partial \mathcal{T}_3$. Furthermore, $\Imag(\tau \circ B \circ \xi ) = \partial \mathcal{T}_3$. By Proposition~\ref{angle function}, this implies that 
    $\{\tau(B(\theta)) : \theta \in \R^{3\times 3}\} = \mathcal{T}_3$, which is the criterion from Equation (\ref{eq:t_implies_enc}).
\end{proof}
This solves the first half of Conjecture 1 in \cite{PRT}. Furthermore by Lemma~\ref{lem:enc_poss_multi}:
\begin{theorem}
    Let $\mathcal{H}_A,\mathcal{H}_B$ be two complex (separable) Hilbert spaces, where one has dimension $3$, and the other one dimension $3\kappa$ for some $\kappa \in \N \cup \{ \infty\}$. Then for any state $\ket{\psi} \in \mathcal{H}_A \otimes \mathcal{H}_B$, the set $\orb_{\LU}(\ket{\psi})$ contains an orthonormal basis.
\end{theorem}

Note that the construction breaks down into multiple steps.\\
(i) Find a group $G$ satisfying Proposition~\ref{prop:group}\\
(ii) Characterize the corresponding set of maximally coherent states (states of the form of Equation (\ref{eq:max_coherent})). All these states satisfy local encodability.\\
(iii) Find a complete set of $\LU$-invariants on $\mathcal{H}$.\\
(iv) Show that the set of maximally coherent states can obtain any value of $\LU$-invariants.\\

For bipartite systems this simplifies to:\\
(i) Check, that $\gcd(d_A,d_B) > 1$.\\
(ii) Use the group and maximally coherent states from example~\ref{group const} (or its simplifications for $d_A = d_B$, or $d_A | d_B$).\\
(iii) Find an angle Function $\xi$ satisfying Proposition~\ref{angle function}.\\

\section{An alternative proof for the $4\times 3$ case }\label{4x3 app}
For $d_A = 4, d_B=3$ the cyclic construction from example~\ref{prod ex} is enough to prove universal local encodability. For a given set of Schmidt coefficients, one must only find angles $(\varphi_{jk})_{jk}$, such that for $A \in \C^{4\times 3}$, $A_{jk} = \frac{e^{i\varphi_{jk}}}{\sqrt{12}}$, the matrix $A^\dagger A$ has eigenvalues $(s_1^2,s_2^2,s_3^2)$. This is indeed always possible.
The key for this construction lies in the combination of the following three statements:\\
(i) The symmetric Schur-Horn theorem (see the first theorem in \cite{SchurHorn}), that states that any set of real eigenvalues (including multiplicity) can be realized by a real symmetric matrix, whose diagonal is majorized by the eigenvalues, and sums to the sum of the eigenvalues. In particular, if the eigenvalues are supposed to be $(s_1^2,s_2^2,s_3^2)$ the diagonal can be chosen as $\left(\frac{1}{3},\frac{1}{3},\frac{1}{3}\right)$.\\
(ii) In the set of correlation matrices
    \begin{equation*}
        \mathcal{R}_n = \{C \in \R^{n\times n} : C^T = C, C\geq 0, \diag(C)=(1,...,1)\},
    \end{equation*}
    those with $\rank(C)\leq 2$ are of the form $C_{jk} = \cos(\varphi_j-\varphi_k)$ for some angles $(\varphi_j)_j$ (see \cite{Correlation} Proposition 1).\\
(iii) In a convex set every element is the midpoint between two boundary points (see \cite{Convex}).
    
Now fix a set of squares of Schmidt-coefficients and consider the matrix $\tilde{C}\in \R^{3\times 3}$ with this spectrum and 

$\diag(\tilde{C})=\left(\frac{1}{3},\frac{1}{3},\frac{1}{3}\right)$. Then $C=3\tilde{C}$ is a correlation matrix, and is the midpoint between two such matrices $C^{(1)}$ and $C^{(2)}$ of rank $\leq2$. Thus
\begin{equation*}
    \tilde{C}_{jk} = \frac{C_{jk}}{3} = \frac{1}{6}\left(C^{(1)}_{jk} +C^{(2)}_{jk}\right) = \frac{1}{6}\left( \Real(e^{i\varphi_j}e^{-i\varphi_k})+\Real(e^{i\theta_j}e^{-i\theta_k}) \right) =(A^\dagger A)_{jk}, \quad A = \frac{1}{\sqrt{12}} \begin{bmatrix} e^{i\varphi_1} & e^{i\varphi_2}&e^{i\varphi_3}\\ e^{-i\varphi_1} & e^{-i\varphi_2}&e^{-i\varphi_3} \\ e^{i\theta_1} & e^{i \theta_2} &e^{i\theta_3} \\
e^{-i\theta_1}  & e^{-i \theta_2} &e^{-i\theta_3}  \end{bmatrix}.
\end{equation*}

This immediately implies:
\begin{theorem}\label{alt 4x3 theorem}
    Universal local encodability is possible for $4\times 3$.
\end{theorem}

\section{Proof of the No-Go Theorem}
\begin{rem}\label{vieta}
    The criterion of Paulwels and Gühne on the dimensions of $\mathcal{H}_j$ in Equation (\ref{ineq 2}) appears to be slightly weaker than the naive bound from Equation (\ref{eq:dim_ineq}) obtained through dimension counting. But they are equivalent, since the edge case
    \begin{equation*}
        \sum_{j=1}^n d_j^2 = (n-1) + \prod_{j=1}^n d_j
    \end{equation*}
    does not occur for $d_j\geq 2$.
\end{rem}
\begin{proof}
    Assume without loss of generality, that $d_1 \leq d_2 \leq ... \leq d_n$. \\
    \underline{$n=2$:} The equation is $d_1^2+ d_2^2 = 1+d_1d_2$ which is equivalent to $(d_1-d_2)^2+d_1d_2 = 1$ and thus impossible, since $d_1d_2 \geq 4$.
    \\
    \underline{$n=3$:} Here the case $d_1 =2$ cannot happen, since this implies $4+d_2^2+d_3^2 = 2+2d_2d_3$ which is equivalent to $(d_2-d_3)^2=-2$. (For $d_1 \geq 3$, the general argument will hold).\\
    
    The proof now follows induction:
    Assume the statement is true for some $n-1 \in \mathbb{N}$. Let $(d_1,...,d_n)\in \N_{\geq 2}^n$ be a solution to the Equation, chosen such that $d_n$ is minimal. Now apply the technique of Vieta root swapping: $d_n$ is a root of the polynomial
    $f(x)=x^2 -\alpha x +\beta$ where $\alpha = \prod\limits_{j=1}^{n-1} d_j$ and $\beta =\sum\limits_{j=1}^{n-1} (d_j^2-1)$. This polynomial has another root $d_n'$, which corresponds to another solution $(d_1,...,d_{n-1},d_n')$ provided $d_n' \in \N_{\geq 2}$.
    By Vieta
    \begin{equation*}
        d_n+d_n'= \alpha \in \mathbb{N}, \quad d_nd_n' = \beta > 0,
    \end{equation*}
    implying $d_n' \in \mathbb{N}$. If $d_n'=1$ then $(d_1,...,d_{n-1})$ is a solution for the $n-1$ case, which is impossible by induction. So $d_n' \in \N_{\geq 2}$. By assumption this implies $d_n' \geq d_n \geq d_{n-1}\geq...\geq d_1$. At the same time
    \begin{equation*}
    \begin{split}
        f(d_{n-1}) = d_{n-1}^2\left(2 - \prod_{j=1}^{n-2} d_j\right)+ \sum_{j=1}^{n-2}d_j^2-(n-1)
        &\leq d_{n-1}^2\left(2 - 2^{n-2}\right)+ (n-2)d_{n-1}^2-(n-1) \\
        &=d_{n-1}^2(n-2^{n-2})-(n-1) <0
    \end{split}
    \end{equation*}
    for $n\geq 4$. For $n=3$ the additional proven identity $d_1 \geq 3$ yields
    \begin{equation*}
        f(d_2) = d_2^2\left(2 - d_1\right)+d_1^2-2 \leq d_{2}^2\left(2 - 3+1\right)-2<0
    \end{equation*}
    Since $f$ is asymptotically positive, this implies that $d_{n-1}$ is strictly in between the two roots $d_n$ and $d_n'$. This contradicts $d_{n-1} \leq d_n \leq d_n'$.
\end{proof}
The same type of argument also shows that $\sum\limits_{j=1}^n d_j^2 = n + \prod\limits_{j=1}^nd_j$ is impossible (which is the edge case in the inequality needed for the existence of a continuum of one-uniform states).

\begin{prop}\label{stab}
    Let $\ket{\psi}$ be a one-uniform state. Then it is regular if and only if its stabilizer Lie-group
    \begin{equation}
        S_{\ket{\psi}}=\left\{ M \in \bigtimes_{j=1}^n \SL(d_j,\C) : M\ket{\psi} = \ket{\psi} \right\}
    \end{equation}
    is discrete (it fulfills $\dim(S_{\ket{\psi}}) = 0$).
\end{prop}
\begin{proof}
    The Lie-algebra corresponding to $S_{\ket{\psi}}$ is given by
    \begin{equation*}
        \mathfrak{s}_{\ket{\psi}} = \left\{M \in \bigoplus_{j=1}^n \mathfrak{sl}(d_j,\C) : M \ket{\psi} = 0\right\}
    \end{equation*}
    where the direct sum of vector spaces is embedded into $L(\mathcal{H})$ via $(x_1,...,x_n) \mapsto x_1 \otimes \mathbbm{1}^{\otimes (n-1)} +...+\mathbbm{1}^{\otimes (n-1)}x_n$.
    Consider now the space of counterexamples to regularity
    \begin{equation*}
        \mathfrak{r}_{\ket{\psi}} = \left\{ X \in \bigoplus_{j=1}^n \mathfrak{su}(d_j) : X \ket{\psi} = 0\right\}.
    \end{equation*}
    Note that $\bigoplus\limits_{j=1}^n \mathfrak{su}(d_j)$ consists precisely out of elements of the form $i\sum\limits_{T \in \mathcal{G}} v_{T} T$ (and the global prefactor $i$ may be ignored).
    For one-uniform states it will be shown that $\mathfrak{r}_{\ket{\psi}} \otimes_{\R} \C = \mathfrak{s}_{\ket{\psi}}$, thus $\mathfrak{s}_{\ket{\psi}}$ is the complexification of $\mathfrak{r}_{\ket{\psi}}$ and
    \begin{equation*}
        \dim_{\R}(\mathfrak{r}_{\ket{\psi}}) = \dim_{\C}(\mathfrak{s}_{\ket{\psi}})=\dim_{\C}(S_{\ket{\psi}}).
    \end{equation*}
    For any discrete $S_{\ket{\psi}}$ this then implies $\mathfrak{r}_{\ket{\psi}}=\{0\}$.\\
    \underline{Proving $\mathfrak{r}_{\ket{\psi}} \otimes_{\R} \C = \mathfrak{s}_{\ket{\psi}}$:}
    \begin{equation*}
    \begin{split}
        \mathfrak{r}_{\ket{\psi}} \otimes_{\R} \C &= \!\left\{ \!X+iY \!:\!X,Y \in \bigoplus_{j=1}^n \mathfrak{su}(d_j) : X \ket{\psi} 
        = Y\ket{\psi}=0\right\} \\
        &=\left\{\! M :M \in \bigoplus_{j=1}^n \mathfrak{sl}(d_j,\C) :  \frac{M-M^\dagger}{2} \ket{\psi} = \frac{M+M^\dagger}{2}\ket{\psi}=0\right\}
    \end{split}
    \end{equation*}
    using the decomposition $M = \frac{M-M^\dagger}{2}+\frac{M+M^\dagger}{2}$ for $M \in\mathfrak{sl}(d,\C)$.
    It remains to show: If $\ket{\psi}$ is one-uniform and $M=X_1 +iX_2$ fulfills $M\ket{\psi} =0$ for $X_1,X_2 \in \bigoplus\limits_{j=1}^n \mathfrak{su}(d_j)$, then $X_1 \ket{\psi} =X_2 \ket{\psi}=0$. Now
    \begin{equation*}
        0=|M\ket{\psi}|^2 = |X_1 \ket{\psi}|^2 + |X_2\ket{\psi}|^2 + i\braket{\psi}{[X_2,X_1] \psi}
    \end{equation*}
    Since $\bigoplus\limits_{j=1}^n \mathfrak{su}(d_j)$ is a Lie-algebra $X \equiv [X_2,X_1] \in\bigoplus\limits_{j=1}^n \mathfrak{su}(d_j) $. It remains to show that for any $X \in \bigoplus\limits_{j=1}^n \mathfrak{su}(d_j)$ it holds that $\braket{\psi}{X\psi}=0$. But this follows from linearity, since this expectation value vanishes on a basis, due to one-uniformity.
\end{proof}
\begin{lemma}\label{density}
    (a) If the Hilbert space dimensions fulfill the inequality (\ref{ineq 2}) (or equivalently violate Inequality (\ref{eq:dim_ineq})), then there is an open and dense set $U$ of (possible unnormlized) vectors $\ket{\psi} \in \mathcal{H}$, such that for every $\ket{\psi} \in U$ it holds that $\dim(S_{\ket{\psi}}) = 0$.
    \\
    (b) Let $G = \bigtimes\limits_{j=1}^n \SL(d_j,\C)$, then the set $\mathscr{O}=\bigcup\limits_{\stackrel{\mathrm{one-uniform} \;\ket{\psi}}{\gamma \in \C \backslash\{0\}} } \orb_{G}(\gamma\ket{\psi})$ is dense in $\mathcal{H}$ if it is nonempty.
\end{lemma}
\begin{proof}
    This condition on the dimension is equivalent to $\Delta(d) > 0$ from \cite{Bryan}. By their Theorem 1.2 it holds for $\Delta(d)>-2$, that $\dim(\mathcal{S}_{\LME}/K) = \Delta(d) \geq 0$, where $\mathcal{S}_{\LME}/K$ is the set of one-uniform states up to local unitary equivalence. In particular this set is nonempty. (The cases of $\Delta(d)\in \{-1,0\}$ are impossible by Remark~\ref{vieta} or \cite{bryan2018existence} Theorem 1.2.)\\
    
    (a) 
    According to Chapter 5 in \cite{Bryan}, there is a subgroup $S \subset G$, and a dense open subset $U \subseteq \mathcal{H}$, such that $S_{\ket{\psi}}$ is conjugate to $S$ (there is a $g\in G$ such that $S_{\ket{\psi}}= gSg^{-1}$) for all $\ket{\psi} \in U$. Furthermore in their notation 
    \begin{equation*}
    \dim(S)=\Delta(d)-\dim(\mathcal{S}_{\LME}/K)=0
    \end{equation*}
    (The first equality is their Equation (1.10) and the second their Theorem 1.2).
    \\
    (b) By \cite{Bryan} the set $\mathscr{O}$ is precisely the set of elements in $\mathcal{H}$ whose orbits are polystable. (This means: The function ${g \mapsto |g\ket{\psi}|}$ has a unique strictly positive minimum. These minima are precisely the one-uniform states up to normalization). Furthermore the union of polystable orbits is dense according to \cite{Slowik} Section 2.1, whenever the set is nonempty. Note, that this density is in the projective space, so that normalization has to be taken care of.
\end{proof}

\begin{theorem}\label{th:impossible}
    If the Hilbert space dimensions fulfill the inequality (\ref{ineq 2}) (or equivalently violate Inequality (\ref{eq:dim_ineq})), then universal local encodability is impossible.
\end{theorem}
\begin{proof}
    According to Proposition~\ref{prop:criteria} it is sufficient to verify, that a regular, one-uniform state $\ket{\psi}$ exists. Consider the two sets $U$ and $\mathscr{O}$ from Lemma~\ref{density}. Since both are dense and $U$ is open, their intersection is nonempty. (Indeed since $\mathscr{O}$ is dense, its complement cannot contain an open set, since it would not be included in its closure.) Thus there exists a one-uniform vector $\ket{\psi}$ and a $g \in G$, such that $S_{g\ket{\psi}}$ is discrete. But $S_{g\ket{\psi}} = gS_{\ket{\psi}}g^{-1}$, so that $S_{\ket{\psi}}$ is discrete as well. Normalizing $\ket{\psi}$ yields a regular one-uniform state. Thus due to Proposition~\ref{stab} the criterion of Proposition~\ref{prop:criteria} is fulfilled.
\end{proof}

\section{Proof of Lemma~\ref{lem:LUdisceps}}\label{ap:POVM}

Here, we prove Lemma~\ref{lem:LUdisceps} of the main text. To that end, recall that we consider, for fixed state $\rho_0 = \ketbra{\psi}{\psi}$, the minimal average loss w.r.t.~POVM measurements 
\begin{align*}
    R_{\rm POVM}^\varepsilon(\rho_0):=\min_{M\in \text{POVM}} R^\varepsilon(M,\rho_0)= \min_{M\in \text{POVM}}\int_{\LU} \text{d}V \int_{\LU} \text{d}\widehat V\,  c_\varepsilon(\widehat V^{\dagger}V) \Tr(\rho_V M_{\widehat V}),
\end{align*}
and likewise $R_{\rm PVM}^\varepsilon$ where we minimize over projective measurements instead. Here, $c_\varepsilon(V)$ denotes the fixed trace-distance tolerance
\begin{align*}
    c_\varepsilon(V) = \begin{cases}1 & \frac12\Vert \rho_0 - V\rho_0 V^\dagger \Vert_1 > \varepsilon, \\ 0 & \text{otherwise.}\end{cases}
\end{align*}
Using the invariance of the Haar measure, we can write
\begin{align}\label{eq:RwithC}
    R^\varepsilon(M,\rho_0) = \int_{\LU} \text{d}\widehat V\,\Tr[\widehat V^\dagger M_{\widehat V} \widehat V C_\varepsilon(\rho_0)]
\end{align}
with the cost operator
\begin{align}\label{eq:costeps}
    C_\varepsilon(\rho_0)= \int_{\LU} \text{d}U\,c_\varepsilon(U) U\rho_0 U^\dagger.
\end{align}
For our choice of $c_\varepsilon$, we can split the integration into two parts. To that end, define the subset of local unitaries mapping to $\varepsilon$-close states,
\begin{align*}
    K_\varepsilon(\rho_0) := \left\{U\in\LU\,:\, \frac12\Vert \rho_0 - U\rho_0 U^\dagger\Vert \leq \varepsilon\right\}
\end{align*}
and its complement $K^c_\varepsilon(\rho_0) := \LU \setminus K_\varepsilon(\rho_0)$ 
Then,
\begin{align}\label{eq:costeps2}
    C_\varepsilon(\rho_0) &= \int_{K^c_\varepsilon(\rho_0)} \text{d}U\, U\rho_0U^\dagger = \int_{\LU} \text{d}U\, U\rho_0U^\dagger -\int_{K_\varepsilon(\rho_0)} \text{d}U\, U\rho_0U^\dagger\nonumber \\
    &= \frac{\id}{D} - \mu_\varepsilon \rho_\varepsilon,
\end{align}
where $\mu_\varepsilon := \int_{K_\varepsilon(\rho_0)} \text{d}U$ is the volume of $K_\varepsilon(\rho_0)$ and $\rho_\varepsilon = \int_{K_\varepsilon(\rho_0)}\text{d}U\,U\rho_0U^\dagger / \mu_\varepsilon$ is the normalized average state within $K_\varepsilon(\rho_0)$.
Before continuing, let us establish two useful properties of norms involving $\rho_0$ and $V\rho_0V^\dagger$ for all $V \in K_\varepsilon(\rho_0)$.
First, knowing that $\Vert \rho_0 - V \rho_0 V^\dagger \Vert_1 \leq 2\varepsilon$, we can bound its operator norm as well. To that end, decompose 
\begin{align*}
    \rho_0 - V\rho_0 V^\dagger = P - N,
\end{align*}
where $P \geq 0$, $N \geq 0$. Tracelessness implies $\Tr(P) = \Tr(N)$, and the trace norm bound means $\Tr(P) + \Tr(N) \leq 2\varepsilon$, thus $\Tr(P) = \Tr(N) \leq \varepsilon$. The operator norm, on the other hand, is given by
\begin{align} \label{eq:operatornormbound}
    \Vert \rho_0 - V\rho_0 V^\dagger \Vert_\infty = \max(\Vert P\Vert_\infty,\Vert N \Vert_\infty) = \max(\Tr(P), \Tr(N)) \leq \varepsilon.
\end{align}
Second, consider the expression
\begin{align*}
    \Tr[(\rho_0 - V\rho_0 V^\dagger)^2] = 2\Tr(\rho_0^2) - 2\Tr(\rho_0 V \rho_0 V^\dagger).
\end{align*}
Using that we choose $\rho_0$ to be pure together with Hölder's inequality, it follows that 
\begin{align} \label{eq:hilbertschmidtbound}
    1-\Tr(\rho_0 V \rho_0 V^\dagger) &= \frac12
  \Tr[(\rho_0 - V\rho_0 V^\dagger)^2] \nonumber \\
  &\leq \frac12 \Vert \rho_0 - V\rho_0 V^\dagger \Vert_1 \Vert \rho_0 - V\rho_0 V^\dagger \Vert_\infty \nonumber \\
  &\leq \varepsilon^2.
\end{align}

With these norm bounds at hand, we can continue by upper bounding the minimal average loss w.r.t.~POVMs. To that end, consider the POVM $M'$ with effects $M'_{\widehat V} := D \widehat V \rho_0 \widehat V^\dagger$, which can immediately be verified to yield a proper POVM. Then,
\begin{align}\label{eq:RPOVMbound}
    R_{\rm POVM}^\varepsilon(M',\rho_0) &\leq R^\varepsilon(M',\rho_0) \overset{\eqref{eq:RwithC}}{=}  D\Tr[\rho_0  C_\varepsilon(\rho_0)] \nonumber \\
    &\overset{\eqref{eq:costeps2}}{=}  1-D\int_{K_\varepsilon(\rho)} \text{d}V\,\Tr[\rho_0 V\rho_0V^\dagger] \nonumber \\
    &\overset{\eqref{eq:hilbertschmidtbound}}{\leq} 1-D\mu_\varepsilon(1-\varepsilon^2)  \leq 1-D\mu_\varepsilon(1-\varepsilon).
\end{align}

Next, we lower bound the minimal average loss w.r.t.~projective measurements. To that end, fix an arbitrary projective measurement $P$ that measures in a fixed orthonormal basis $\{\ket{e_a}\}$ and reports, upon receiving outcome $a$, the local unitary $\widehat V_a$. Thus,
\begin{align}\label{eq:RPVMbound}
    R_{\rm PVM}^\varepsilon(\rho_0) &\overset{\eqref{eq:RwithC}}{=} \min_{\{\ket{e_a}\},\{\widehat V_a\}} \sum_{a=1}^D \Tr[\widehat V_a^\dagger \ketbra{e_a}{e_a} \widehat V_a C_\varepsilon(\rho_0)] \nonumber \\
    &\overset{\eqref{eq:costeps2}}{=}  \min_{\{\ket{e_a}\},\{\widehat V_a\}} \left(1-\mu_\varepsilon \sum_{a=1}^D \braket{e_a}{\widehat V_a \rho_\varepsilon \widehat V_a^\dagger| e_a}\right).
\end{align}
Consider the second term. It can be bounded via
\begin{align*}
    \sum_{a=1}^D \braket{e_a}{\widehat V_a \rho_\varepsilon \widehat V_a^\dagger| e_a} &= \sum_{a=1}^D \braket{e_a}{\widehat V_a (\rho_\varepsilon - \rho_0)  \widehat V_a^\dagger| e_a} + \sum_{a=1}^D \braket{e_a}{\widehat V_a\rho_0  \widehat V_a^\dagger| e_a} \\
    &\leq D\Vert\rho_\varepsilon - \rho_0\Vert_\infty + \sum_{a=1}^D |\braket{e_a}{\widehat V_a|\psi}|^2
\end{align*}
Convexity of the operator norm yields, together with Eq.~\eqref{eq:operatornormbound}, $\Vert \rho_\varepsilon - \rho_0\Vert_\infty \leq \varepsilon$. Plugging this into Eq.~\eqref{eq:RPVMbound} yields
\begin{align}\label{eq:RPVMbound2}
    R_{\rm PVM}^\varepsilon(\rho_0) \geq 1-\mu_\varepsilon\left (D\varepsilon + \max_{\{\ket{e_a}\},\{\widehat V_a\}} \sum_{a=1}^D  |\braket{e_a}{\widehat V_a|\psi}|^2\right).
\end{align}
The maximization in this term reaches its maximum value of $D$  if and only if  $\ket{\psi}$ is locally encodable: We can then choose $\widehat V_a$ as the local encoders and $\ket{e_a}$ as the encoded basis vector, yielding the maximal value of $1$ in each term of the sum. If, however, $\ket{\psi}$ is not locally encodable, due to compactness of the domain of maximization, then at least one of the terms remains smaller than $1$. 
Thus, we can quantify the local encodability of $\ket{\psi}$ by writing 
\begin{align*}
    \max_{\{\ket{e_a}\},\{\widehat V_a\}} \sum_{a=1}^D  |\braket{e_a}{\widehat V_a|\psi}|^2 = D(1-\gamma)
\end{align*}
for some $\gamma \geq 0$, and $\gamma = 0$ if and only if $\ket{\psi}$ is locally encodable.

Thus, Eq.~\eqref{eq:RPVMbound2} yields
\begin{align*}
    R_{\rm PVM}^\varepsilon(\rho_0) \geq 1-D\mu_\varepsilon(1-\gamma+\varepsilon).
\end{align*}
Comparing this lower bound for projective measurements with the upper bound in Eq.~\eqref{eq:RPOVMbound} shows that choosing any $\varepsilon < \gamma/2$ implies
\begin{align*}
    R_{\rm POVM}^\varepsilon(\rho_0) < R_{\rm PVM}^\varepsilon(\rho_0),
\end{align*}
proving the claim.

\end{document}